\documentclass[letter,11pt]{article}
\pdfoutput=1
\usepackage{jheppub} 
\usepackage{graphicx}

\usepackage{graphicx}
\usepackage{dcolumn}
\usepackage{bm}
\usepackage{subfigure}
\usepackage{verbatim}
\usepackage{color}
\usepackage{lineno}

\begin{document}

\preprint{\small FERMILAB-PUB-15-079-E}

\title{Dark Matter Directionality Revisited with a High Pressure Xenon Gas Detector 
}

\author[a]{Gopolang Mohlabeng,}
\affiliation[a]{Department of Physics and Astronomy, University of Kansas, Lawrence, KS 66045, USA.}

\author[a]{Kyoungchul Kong,}

\author[b]{Jin Li,}
\affiliation[b]{Center for Underground Physics, Institute for Basic Science (IBS), Daejon 305-811, Korea}

\author[c]{Adam Para,}
\affiliation[c]{Fermi National Accelerator Laboratory, Batavia, IL 60510, USA. }

\author[c]{Jonghee Yoo}

\emailAdd{gopolang.mohlabeng@ku.edu}
\emailAdd{kckong@ku.edu}
\emailAdd{jinlee@ibs.re.kr}
\emailAdd{para@fnal.gov}
\emailAdd{yoo@fnal.gov} 

\date{\today}

\abstract{
An observation of the anisotropy of dark matter interactions in a direction-sensitive detector would provide decisive evidence for the discovery of galactic dark matter. Directional information would also provide a crucial input to understanding its distribution in the local Universe. Most of the existing directional dark matter detectors utilize particle tracking methods in a low-pressure gas time projection chamber. These low pressure detectors require excessively large volumes in order to be competitive in the search for physics beyond the current limit. In order to avoid these volume limitations, we consider a novel proposal, which exploits a columnar recombination effect in a high-pressure gas time projection chamber. The ratio of scintillation to ionization signals observed in the detector carries the angular information of the particle interactions.

In this paper, we investigate the sensitivity of a future directional detector focused on the proposed high-pressure Xenon gas time projection chamber. We study the prospect of detecting an anisotropy in the dark matter velocity distribution. We find that tens of events are needed to exclude an isotropic distribution of dark matter interactions at 95\% confidence level in the most optimistic case with head-to-tail information. However, one needs at least 10-20 times more events without head-to-tail information for light dark matter below $\sim$50 GeV. For an intermediate mass range, we find it challenging to observe an anisotropy of the dark matter distribution. Our results also show that the directional information significantly improves precision measurements of dark matter mass and the elastic scattering cross section for a heavy dark matter.}

\maketitle

\section{Introduction}
\par The Standard Model (SM) of elementary particle physics has been astonishingly successful in explaining much of the presently available experimental data. However, it still leaves open a number of outstanding fundamental questions whose answers are expected to emerge in a more general theoretical framework. One of the major motivations for pursuing new physics beyond the SM is the `dark matter puzzle', which finds no explanation within the Standard Model. From the accumulated experimental data, we now know that ordinary matter comprises only about 4.9\% of the Universe. The remaining 95.1\% is divided between a mysterious form of matter called `dark matter' (26.8\%) and an even more perplexing entity called `dark energy' (68.3\%) \cite{Ade:2013ktc}.

Naturally, discovering dark matter (DM) and measuring its properties has become central to the fields of particle physics, astrophysics and cosmology. The diversity of possible dark matter candidates requires a well-balanced program based on direct detection experiments, indirect detection experiments, collider experiments and astrophysical probes sensitive to the non-gravitational interactions of dark matter. Vast experimental and technological progress in the coming decade will put the most promising ideas to the test \cite{Arrenberg:2013rzp}.

In the standard scenario of a WIMP (weakly interacting massive particle), direct detection experiments record the nuclear recoil energy spectra produced when a dark matter particle scatters off a target nucleus. The expected nuclear recoil energy falls exponentially and such events, with an energy of typically not more than a few tens of keV, lie well within the range of abundant backgrounds due to radioactivity and other cosmogenic backgrounds. Despite these challenges, experimental limits on the interaction cross-section versus WIMP mass have been steadily improved. Yet, there exists no widely accepted evidence of their presence on Earth. Firm evidence of directionality relative to a WIMP wind would be the most robust signature of the WIMP nature of dark matter, and is an essential step to a claim of discovery. If any of the direct detection experiments observe evidence for nuclear recoils that cannot be explained with known processes, then the search for directionality in such nuclear recoils will be of foremost interest. Given that dark matter signatures exhibit an exponentially falling energy spectrum, and an annual modulation of interaction rates which can be easily mimicked by any activity of seasonal variations, or cosmogenic backgrounds, a discovery claim of dark matter would have to be followed by other exceptional evidence \cite{Gondolo:2002np, Copi:2000tv, Bernabei:2010mq, Ahlen:2009ev}.

A powerful signature of dark matter would be the detection of a significant spatial anisotropy in the angular distribution of such nuclear recoils consistent with the standard model of a non-co-rotating WIMP halo. Earth's position in the galactic arm provides a boost of $\sim$230 km/s, comparable to the quasi-virial velocity $\sim$220 km/s of gravitationally captured WIMPs. Dark matter interactions would produce a large forward-backward asymmetry in the angular distribution of nuclear recoils. An importance of the directional information for an incontestable claim of discovery and its role in the additional rejection of terrestrial backgrounds is widely appreciated~\cite{Spergel:1987kx, Mayet:2013mpa, Billard:2011zj, Mayet:2012qw,Bozorgnia:2012eg,Green:2006cb, Lee:2014cpa,Alves:2012ay,Kavanagh:2015aqa,Green:2010gw,Drukier:2012hj}. 

Currently most attempts at directional detection have focused on low-pressure gas time projection chambers (TPCs), in order to provide the 3-D track reconstruction and energy resolution needed to identify low energy nuclear recoils~\cite{Santos:2013oua, Burgos:2007gv, Daw:2013waa, Miuchi:2007ga,Miuchi:2010hn, Sciolla:2008ku,Billard:2012bk, Billard:2013cxa, Battat:2014van, Lee:2012pf,Morgan:2012sv,Green:2007at}. In all low-pressure TPC detectors, strong tension exists between the desire to use a very low gas density so that nuclear recoil tracks are long enough to be imaged with adequate clarity, and the desire to increase the gas density so that greater sensitivity can be realized. The Diffusion of the ionization image during drift,  limits the drift length and the avalanche amplification noise and/or photon detection quantum efficiency degrades the quality of the track information. In most cases, the total mass per detector in these approaches is less than a kg due to the limited scalability of the low pressure detector. Thus progressing to ton-scale masses would imply a very large and impractical number of separate devices.

Recently, a novel approach has been proposed to confront the challenges of the directional sensitivity for nuclear recoils with active masses approaching the ton-scale. The detector concept is based on a high-pressure Xenon gas TPC with an electroluminescent gain stage which utilizes the `columnar recombination' (CR) process, leading to a potential directional sensitivity of nuclear recoils \cite{Nygren:2013nda}. If this conceptual idea and the related detector technology can be demonstrated, it would revolutionize dark matter experiments\cite{Billard:2014ewa}. Unlike low pressure gas tracking detectors, a ton-scale high-pressure gas dark matter detector would be more practical.

In this paper we examine the sensitivity of the proposed high-pressure Xenon TPC directional dark matter detector. We pay special attention to the capability of measuring head-to-tail information and distinguishing the incoming and outgoing direction of the recoiled nucleus. We study the prospect of detecting an anisotropy in the dark matter velocity distribution. We begin our discussion with a short review on directional dark matter detection in Section \ref{sec:review}. We devote Section \ref{sec:analysis} to a detailed analysis involving a high-pressure Xenon gas detector. We discuss more on columnar recombination, set up our numerical study (Section \ref{sec:setup}) and examine how much improvement can be made on the measurements of mass and cross-section (Section \ref{sec:param}). We then investigate the angular distributions and anisotropy of the dark matter distribution in Section \ref{sec:anglular} and Section \ref{sec:cr}. Section \ref{sec:conclusion} is reserved for discussion.

\section{A Brief Review on Dark Matter Directionality\label{sec:review}}
The motion of the solar system relative to the Galactic WIMP halo provides a distinctive signal for WIMP detection. This circular orbit of our solar system around the galactic center results in a very strong forward-backward asymmetry in the angular distribution of nuclear recoils produced in WIMP events. The differential nuclear recoil rates as a function of both recoil energy and recoil angle have been extensively studied in literature. Recoil rates including the angular distribution of events were first discussed in Ref. \cite{Spergel:1987kx} and then further developed in Refs. \cite{Copi:2000tv, Gondolo:2002np,Lewin:1995rx}. In this study we adopt the formalisms used in Refs. \cite{Copi:2000tv,Gondolo:2002np,Bozorgnia:2011vc} and only provide a short review in our paper.

\begin{figure}[t]
\centering
\includegraphics[scale=0.65]{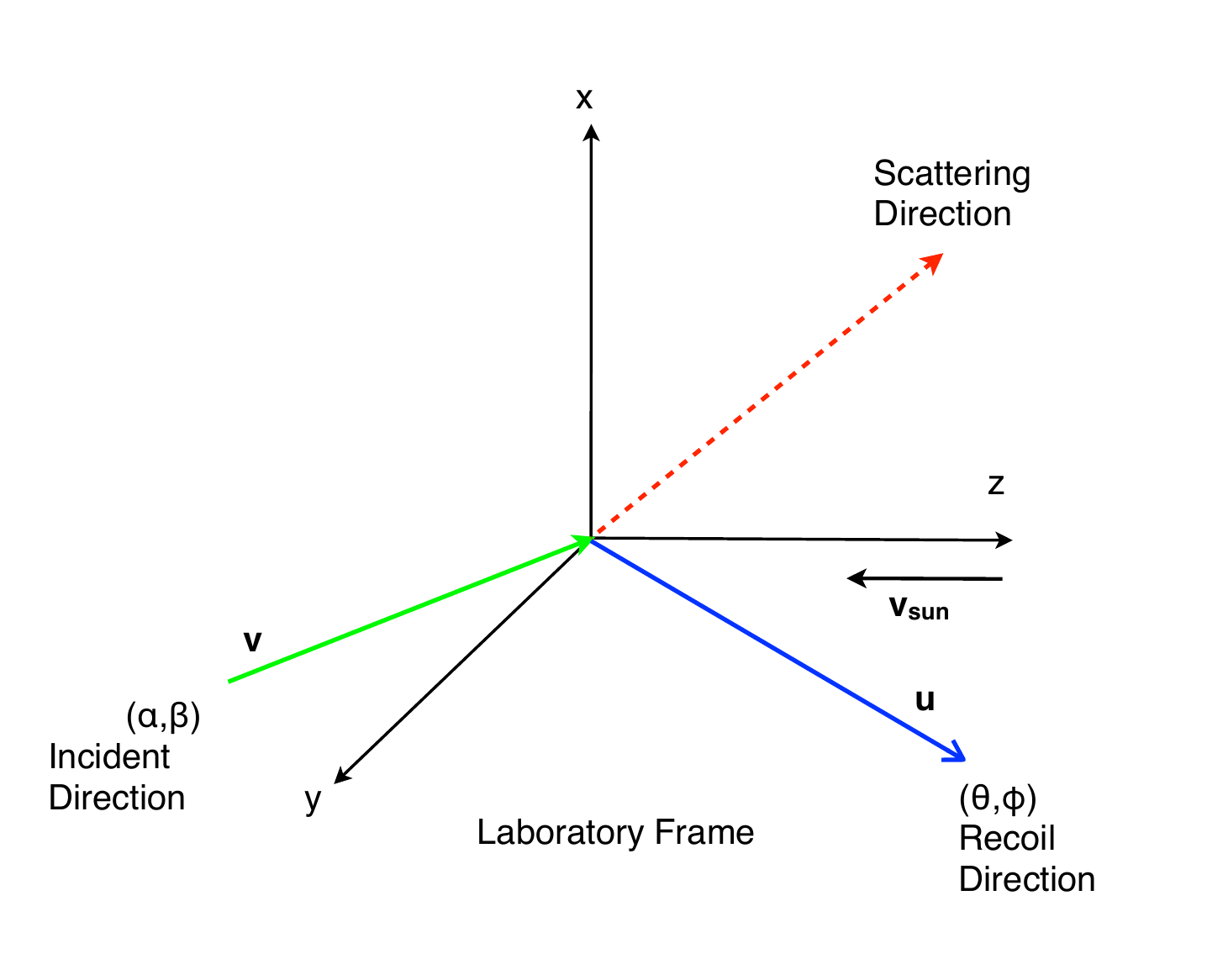}
\caption{Geometry of a WIMP scattering off a target nucleus in the detector. The WIMP is incident at an angle $(\alpha, \beta)$ relative to the ${\bf z}$ axis. The nucleus recoils in the $(\theta, \phi)$ direction.}
\label{wimpd}
\end{figure}

Let us consider a WIMP particle of mass $M_{\chi}$, incident at velocity ${\bf v} = v ( \sin \alpha \cos \beta {\bf  \hat x} + \sin \alpha \sin \beta {\bf  \hat y} + \cos \alpha {\bf \hat z})$ in the detector, as illustrated in Figure \ref{wimpd}. After interaction with a WIMP, the target nucleus recoils with some velocity ${\bf u} = u(\sin \theta \cos \phi {\bf \hat x} + \sin \theta \sin \phi {\bf \hat y} + \cos \theta {\bf \hat z})$ and momentum {\bf q} at a direction $(\theta, \phi)$. 
The rate at which this nucleus recoils per unit recoil energy per unit recoil angle is given as follows:
\begin{eqnarray}
  \frac{d^{2}R}{dE_{R} d\Omega_{(\theta,\phi)}} = \frac{N_{0} \rho_{0} \sigma_{WN}}{\pi \, A \, r M_{\chi}^{2}} F^{2}(q) \int \delta(v \cos \theta - \frac{q}{2 \mu_N}) f({\bf v}) d^{3}v \, , 
\label{doublerate}
\end{eqnarray}
where $\rho_{0}$ is the dark matter halo density in our local part of the galaxy, $\sigma_{WN}$ is the WIMP-nucleus elastic scattering cross-section, and $F(q)$ is the nuclear elastic scattering form factor. Assuming that the nucleus can be approximated to be a sphere with uniform density, the form factor is the Fourier transform of the nuclear density. This gives us the Helm form factor $F(q) = \frac{3[sin(qr_{n})-qr_{n}cos(qr_{n})]}{(qr_{n})^{3}}e^{-(qs)^{2}/2}$, where $r_n$ is an effective nuclear radius and $q = \sqrt{ 2 M_{N} E_{R} }$  is the recoil momentum of the nucleus.  $E_{R}$ is the recoil energy of the nucleus and  $r = 4 M_{N} M_{\chi}/(M_{N} + M_{\chi})^{2}$  is a kinematic factor \cite{Lewin:1995rx}.
Angle $\theta$ is the recoil angle which determines the direction between the recoiling nucleus and the initial WIMP trajectory and $f({\bf v}) $ represents the velocity distribution of WIMPs in the galactic halo.
We call Eq. \ref{doublerate} the double differential recoil rate. $M_{N} = 0.932 A$ GeV is the target mass, with $A$ the atomic mass number of the target atom in atomic mass units (AMU). The factor 0.932 is the value of AMU in GeV and $\mu_N = M_{N} M_{\chi}/(M_{N} + M_{\chi})$ is the reduced mass of the WIMP-Nucleus system. 

The double differential recoil rate can be defined in a simpler mathematical form by adopting the following mathematical convention \cite{Gondolo:2002np}: 
\begin{equation}
{\hat f(v_{q}, \bf \hat q)} = \int \delta({\bf v}.{\bf \hat q} - v_{q}) f({\bf v}) d^{3} v \, ,
\label{radon}
\end{equation}
where $v_{q}$ is the minimum velocity a WIMP must have to impart a recoil momentum $q$ to the nucleus, or equivalently to deposit an energy $E_R = \frac{q^2}{2 M_N}$, ${\bf \hat q}$ is the recoil momentum direction and {\bf v} is the velocity of a WIMP particle in the halo. 
Eq. (\ref{radon}) is the definition of a three-dimensional Radon transformation, which represents the velocity distribution for a stationary detector in the galactic frame. For an observer moving with velocity ${\bf V_{lab}}$ in the galactic frame, this is the velocity of the observer in the galactic frame. It is related to the velocity of the WIMP in the lab frame, ${\bf v_{lab}}$ and in the galactic frame ${\bf v_{gal}}$ by a Galilean transformation ${\bf v_{lab} = \bf v_{gal} - \bf V_{lab}}$. The properties of the Radon transformation for a pure translation $\bf V_{lab}$ \cite{Gondolo:2002np} imply,
\begin{equation}
 {\hat f_{\rm lab}(v_{q}, \bf \hat q)} =  {\hat f_{\rm gal}(v_{q} + \bf V_{lab}.\bf \hat q , \bf \hat q)} \, . \\
 \label{galradon}
\end{equation}

For our study we assume a Maxwell-Boltzmann velocity distribution, truncated at the escape velocity $v_{esc}$ of the WIMPs ;
\begin{equation}
  f_{M}(v) =  \frac{1}{k_{esc} \pi^{3/2} v_{0}^{3}} \exp \left [-\frac{\mid {\bf v} \mid^{2}}{v_{0}^{2}} \right ] \, ,
  \label{maxwell}
\end{equation}
for $v \, \textless \,  v_{esc}$ and $ f_{M}(v) = 0$ otherwise. $k_{esc}$ is a normalization factor which is obtained by integrating the velocity distribution in the galactic frame from 0 to $v_{esc}$. For the velocity distribution Eq. (\ref{maxwell}), the radon transform becomes, 
\begin{equation}
  \hat{f}_{M}(v_{q}, {\bf \hat{q}}) =  \frac{1}{k_{esc} \pi^{1/2} v_{0}} (\exp \left [-\frac{ (v_{q} + {\bf \hat{q}} \cdot  \bf V_{lab})^{2}}{v_{0}^{2}} \right ] - \exp \left [-\frac{v_{esc}^{2}}{v_{0}^{2}} \right ] )\, .
  \label{maxwellradon}
\end{equation}
Finally for a detector on earth moving through the galaxy with velocity ${\bf v_{E}}$ in the direction of Cygnus X-2, ${\bf V_{lab}} = {\bf v_{E}}$ and ${\bf \hat{q}} \cdot {\bf V_{lab}} = - v_{E} \cos \theta $. We can combine Eqs. (\ref{doublerate}) and (\ref{maxwellradon}) to obtain
\begin{eqnarray}
  \frac{d^{2}R}{dE_{R} d\Omega_{(\theta, \phi)}} = \frac{N_0 \, \rho_{0} \, \sigma_{WN}}{\pi^{3/2} A \, r \, v_{0} \, M_{\chi}^{2}} \frac{F^{2}(E_{R}) }{k_{esc}} 
  \left (\exp \left [-\frac{(v_{E} \cos \theta - v_{min})^{2}}{v_{0}^{2}} \right ] - \exp \left [-\frac{v_{esc}^{2}}{v_{0}^{2}} \right ] \right ) \, ,
\label{doubleratereq}
\end{eqnarray}
where $N_0$ is the Avogadro's number and $v_{q} = v_{min} = \sqrt{E_{R}/E_{0} r} \, v_{0}$, with $E_{0} = \frac{1}{2}  M_{\chi} v_{0}^{2}$ the most probable kinetic energy of the WIMPs. 
We choose the most probable WIMP velocity $v_{0}= 230 \, km/s$ and the escape velocity of the WIMPs from the galactic halo, $v_{esc} = 600 \, km/s$ \cite{Lewin:1995rx}. $v_{E}$ is calculated in the appendix of Ref. \cite{Lewin:1995rx} and includes the velocity of the Earth with respect to the Sun, the proper motion of the Sun and the velocity of the solar system with respect to the galactic center.
The WIMP-Nucleus cross section is defined as $\sigma_{WN}= \frac{4}{\pi} \mu_N^2 \left (  f_p Z + (A - Z ) f_n\right )^2$, 
where $f_{p} = \sqrt{ \frac{\pi}{4} \sigma_{W p} \frac{1}{\mu_{p}^2}}$ and similarly for $f_{n}$,  
with $f_{p}$ and $f_{n}$ the WIMP-proton and WIMP-neutron couplings respectively. 
In the case where $f_{p} \sim f_{n}$ (which we assume),  we obtain $\sigma_{WN} = \frac{\mu_N^2}{\mu_p^2}\sigma_0 A^2$ with  
$\sigma_0 = \sigma_{Wn} = \sigma_{Wp}$, the WIMP-nucleon cross-section and $\mu_p \approx \mu_n$ \cite{Lewin:1995rx}.
As the truncated Maxwell-Boltzmann velocity distribution exhibits a rotational symmetry (along $\phi$), Eq. (\ref{doubleratereq}) is only dependent on the polar angle, $\theta$.

\section{Dark Matter Directionality with a High-Pressure Xenon Gas Detector \label{sec:analysis}}

In this paper we focus on two aspects of directional detection, ``parameter estimation'' and ``measurement of anisotropy'', 
with emphasis on a high-pressure Xenon gas detector.
As such a detector does not currently exist and only a concept is discussed \cite{Nygren:2013nda}, 
it is uncertain what features would be appropriate to consider.
Therefore we assume certain important detector parameters for our study. 
We first introduce four different types of detectors (Section \ref{sec:setup}) for discussion and 
compare their performance in the parameter estimation of the WIMP mass and WIMP-nucleon cross section (Section \ref{sec:param}). 
Further information on the angular distributions and an anisotropy in the dark matter flow are presented in Section \ref{sec:anglular} for 
those different detectors. 
Although they are generic dark matter detectors without details of a particular detector concept, 
results are still relevant to grasp potential performance of a high-pressure Xenon gas detector and how it would compare as to non-directional detectors. 
Finally in Section \ref{sec:cr} we study anisotropy of the WIMP velocity distribution using a high pressure gas TPC. 
Throughout our studies, we include detector resolution with relevant energy threshold cuts.

\subsection{Columnar Recombination and the Numerical Set Up\label{sec:setup}}

The importance of directional information in dark matter experiments has been recognized for a long time~\cite{Ahlen:2009ev, Daw:2013waa, Burgos:2008jm, Burgos:2007gv, Dujmic:2007bd, Miuchi:2007ga, Tanimori:2003xs, Santos:2007ga}. The short range of the low energy nuclear recoils is an obvious experimental challenge. Therefore a low pressure gas TPC is a natural experimental choice to extend the observable length of the recoil trajectory up to macroscopic dimensions thus enabling the determination of the spatial direction of the recoil track. Unfortunately about 1/10 bar of low pressure gas limits the practically attainable mass of the detector. Hence the technique is currently only applicable in case of relatively large interaction cross sections.

Recently a conceptual dark matter detector that utilizes columnar recombination (CR)~\cite{Jaffe:1913} has been proposed by D. Nygren \cite{Nygren:2013nda} as a possible technique for the determination of the recoil direction in massive detectors, up to several tons. The detector exploits CR within an ensemble of ions and electrons generated by the nuclear recoil. The CR process occurs when the detection of a highly ionizing track and an externally applied electric field coincide, such that the external field drives the ionization electrons to drift in the vicinity of the ion column. Electrons traveling at distances close to the ions and below the Onsager radius undergo electron-ion recombination with an emission of characteristic photons. Conversely, recombination is much less likely if the particle track and the electric field are perpendicular. When the particle track and E-field coincide maximum CR is expected as opposed to when they are parallel to each other. The amount of CR can give us an estimate of the relative angle between the track and the E-field, thus a measurement of the nuclear recoil angle. Hence, the directional information of nuclear recoils might be obtainable in a high-pressure gas detector.  
 
This preliminary detector concept explores the possibility of utilizing a special Penning mixture in the Xenon gas which will convert the energy harbored in primary excitations to ionization. In addition to the uniform drift electric field and a charge collection plane it would be equipped with internal reflectors and photodetectors allowing for a highly efficient collection of light from the entire volume of the detector. After the initial interaction, the electrons drift along the field direction towards the collection plane. The electrons may undergo recombination, with the emission of characteristic photons. The number of the emitted photons and thus the size of the light signal $S$, will depend on the angle between the recoil track and the drift field direction, therefore it is suggested that the division of a total signal, $S+I$, into its components $S$ and ionization $I$ should depend on the angle $\theta_{L} $ between the recoil track direction and the direction of the electric field $\vec{E}$ in the TPC.

Practical implementation of such a concept awaits experimental demonstration \cite{Billard:2014ewa, Grothaus:2014hja}. In particular the head-to-tail capabilities of the detector are of great importance. The direction of recoil is determined through $\cos \theta_{L}  =f( \frac{ S}{S+I})$ therefore its values are limited to be positive. If the columnar recombination is forward-backward symmetric it will allow for the determination of $\left\vert \cos \theta_{L} \right\vert $, otherwise the mapping of $S/(S+I)$ onto $\left\vert \cos \theta_{L} \right\vert $ may be multi-valued, but there will be some region around  $\cos  \theta_{L}  = 1$ (most likely) or $\cos \theta_{L}  = -1$ with characteristically higher values of $S/(S+I)$. In the latter case one will be able to classify all events into {\it two angular bins}: larger or smaller than some $\cos \theta_{0}$, where a possible value of  $\theta_{0}$  must be established experimentally. 

We compare the physics potential of various classes of detectors with different capabilities of the directional measurement for studies of the dark matter interaction with a cross section of $\sigma_{W n} = 5 \times 10^{-11}$ pb $= 5 \times 10^{-47} cm^{2}$. We also assume that the forthcoming generation of experiments will focus on the demonstration of the galactic origin of the observed signal (directionality) and on the determination of the properties of the dark matter (interaction cross section and mass). Our analysis is restricted to dark matter signal events assuming zero-backgrounds, 
which is a good estimate for a 4 keV threshold cut \footnote{Very little neutrino background is expected for recoil energies above 4 keV  \cite{Billard:2013qya}, {\it e.g.,} about 0.5 neutrino events for a Xenon detector with 10 ton-year.}.

To be specific, we consider the following progression of possible detectors:    
\begin{itemize}
\item a detector with no directional capabilities
\item an `ideal' detector capable of measuring the recoil angle in the range $-1 < \cos \theta < 1$  , with some characteristic resolution 
(we refer to this case as `head-to-tail')
\item a `symmetric columnar recombination' detector, thus capable of the determination of  $\left\vert \cos \theta_{L} \right\vert $, 
with some characteristic resolution (we refer to this case as `no head-to-tail', it is also known as a `folded' directional rate \cite{Alenazi:2007sy}. 
See also Ref. \cite{Copi:2005ya} for related studies.)
\item an `asymmetric columnar recombination' detector, thus capable of classifying the events in two angular bins 
\end{itemize}

In addition we compare the capabilities of directional detectors constructed on a movable system that maintains the orientation of the detector's electric field in the galactic frame -- thus rotating in the Earth coordinates. We use the direction of the Earth's motion as our reference direction, and we define the direction of dark matter flow as our forward direction, which is opposite to Cygnus. We call this the `parallel' case, when $\vec E$ is aligned with our forward direction and  `perpendicular', when $\vec E$ is perpendicular \cite{Cao:2014gns}. We define the corresponding angle between the electric field and the recoil direction as $\theta_L = \theta_\parallel$  and $\theta_L = \theta_\perp$, respectively. This set up conveniently identifies $\theta_L = \theta_\parallel$ as the recoil angle $\theta$ in Eq. (\ref{doubleratereq}), for the parallel case.
\begin{figure}[t]
\centering
\centerline{
 \includegraphics[scale=0.715]{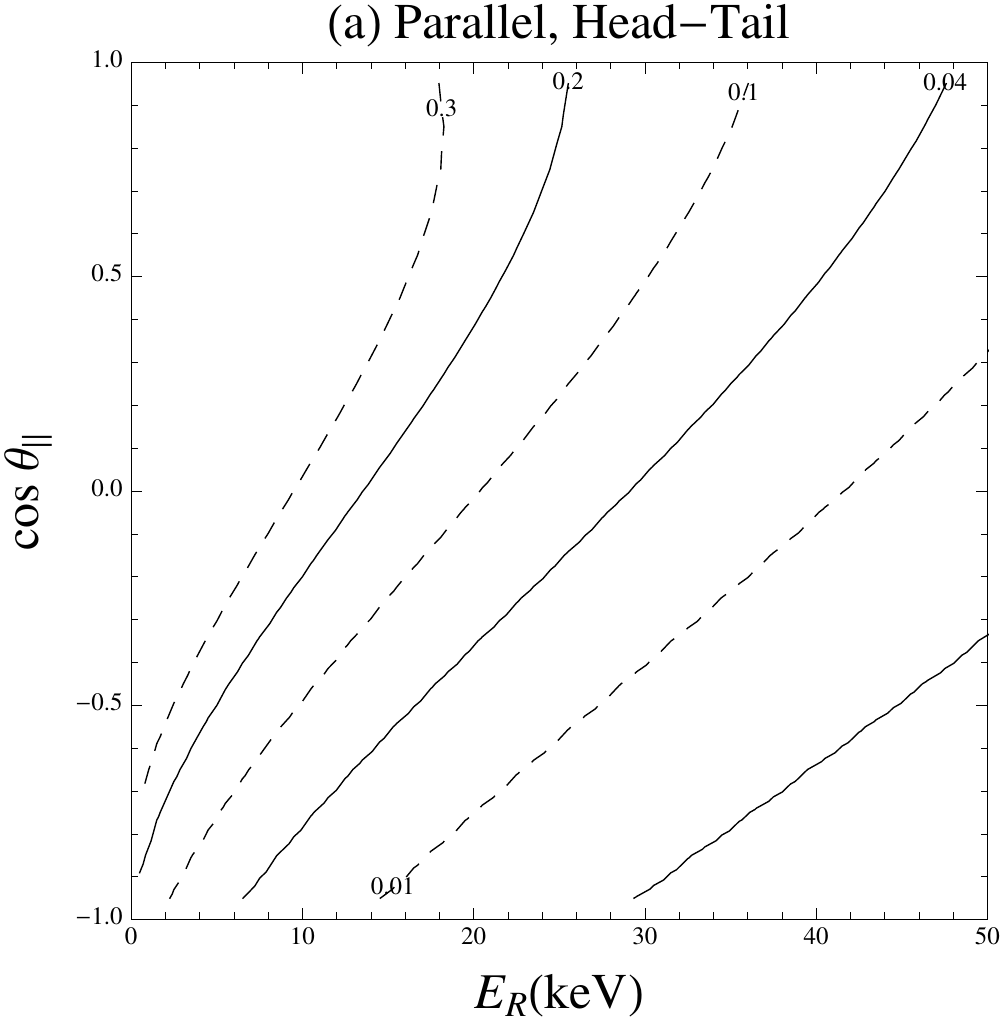}\hspace{0.4cm}
 \includegraphics[scale=0.7]{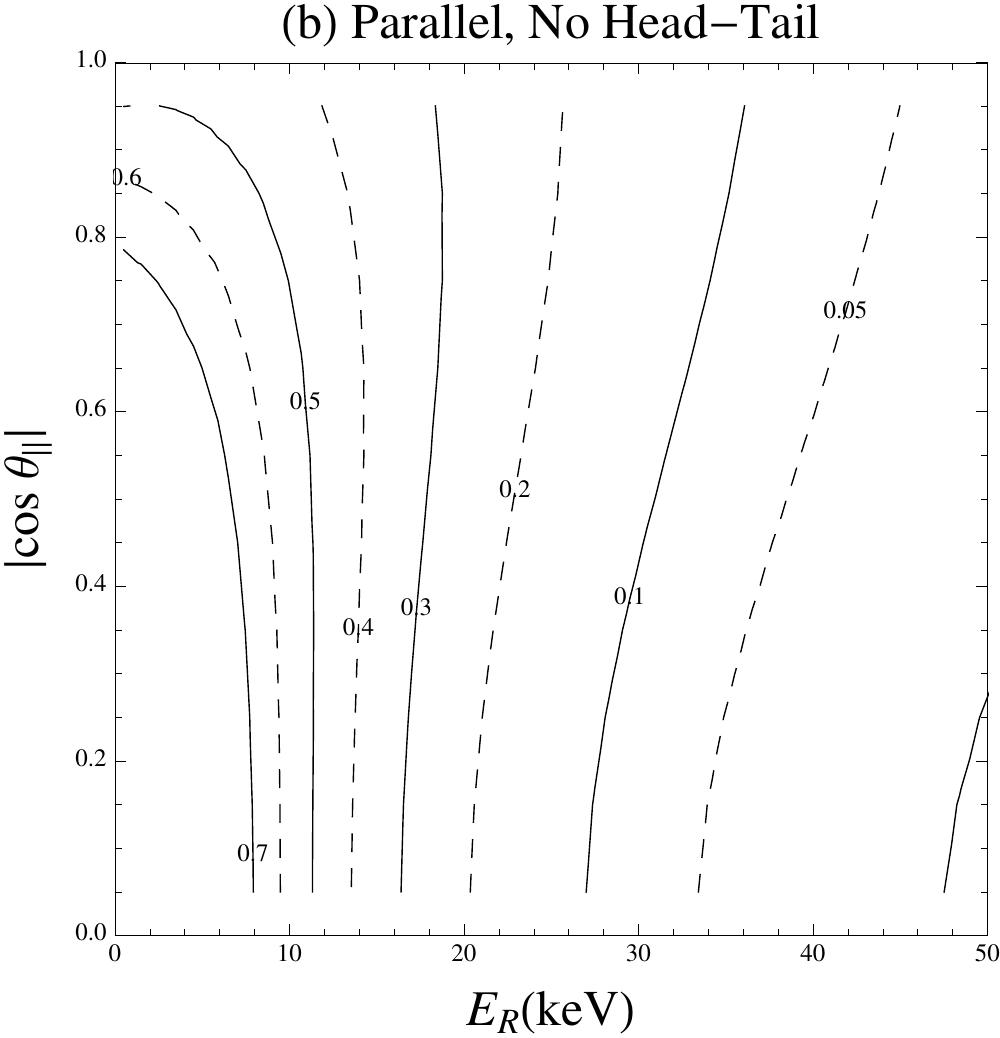}  } 
 \vspace{0.3cm}
 \centerline{
 \includegraphics[scale=0.71]{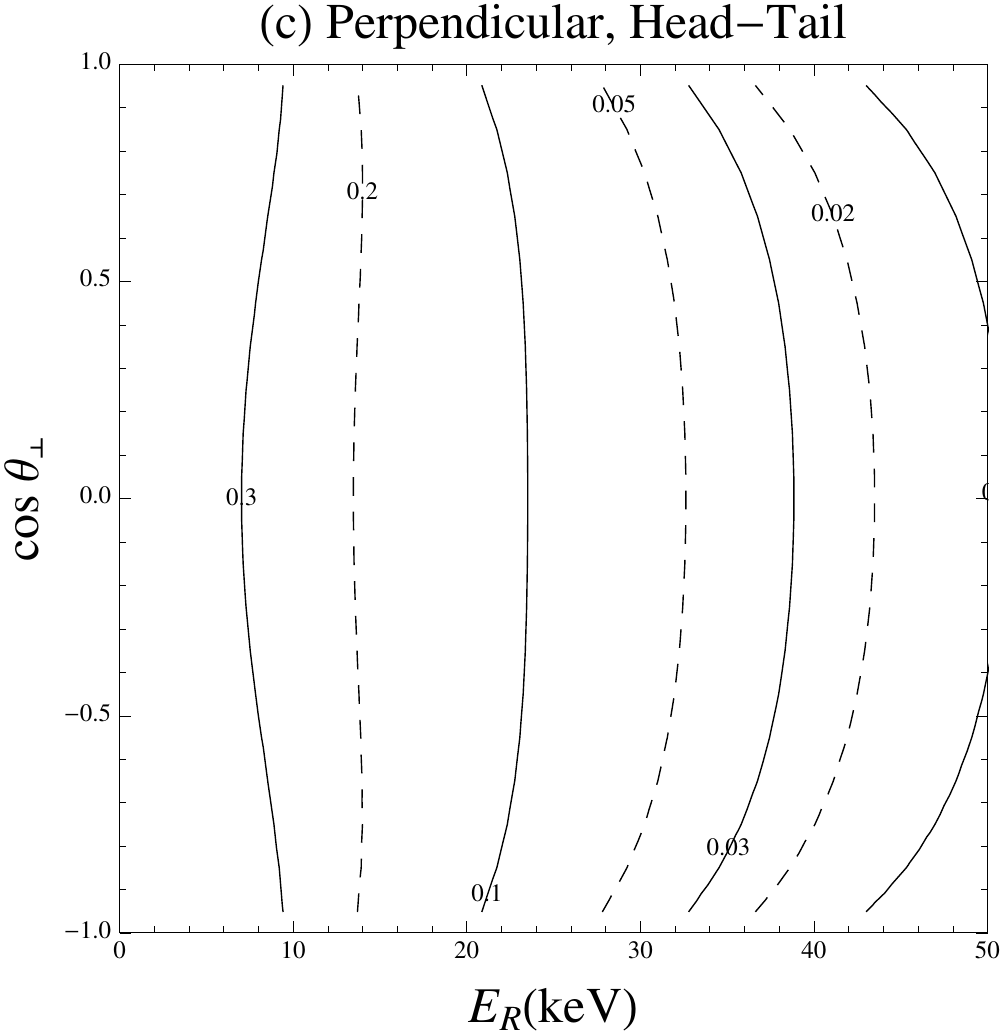}\hspace{0.4cm}
 \includegraphics[scale=0.705]{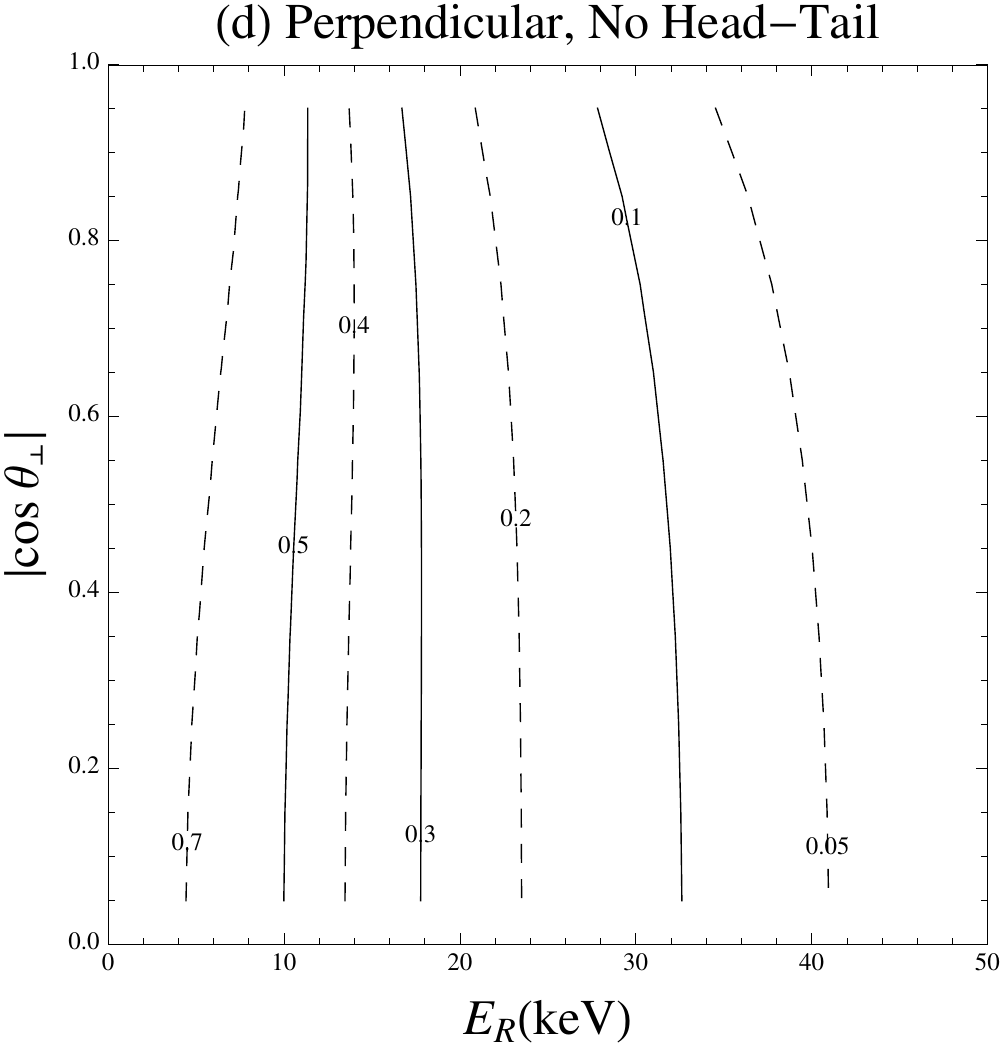} }
\caption{Contours of constant number of events in a two dimensional space for a 10 ton-year Xenon detector. Events are normalized to a case with an 80 GeV dark matter particle and a WIMP-nucleon cross-section of $5\times 10^{-11}$ pb. 
We consider a detector with 
(a) head-to-tail capability and a parallel electric field \big ($  \frac{d^2 N}{ d E_R d \cos \theta_\parallel }$\big), 
(b) no head-to-tail capability with a parallel electric field \big ($  \frac{d^2 N}{ d E_R d | \cos \theta_\parallel | } $\big ), 
(c) head-to-tail capability with a perpendicular electric field \big ($ \frac{d^2 N}{ d E_R d \cos \theta_\perp    }$\big ) and 
(d) no head-to-tail with a perpendicular electric field \big ($\frac{d^2 N}{ d E_R d | \cos \theta_\perp |} $\big ). }
\label{corrhtnht}
\end{figure}

\begin{figure}[t]
\centerline{
\includegraphics[scale=0.5]{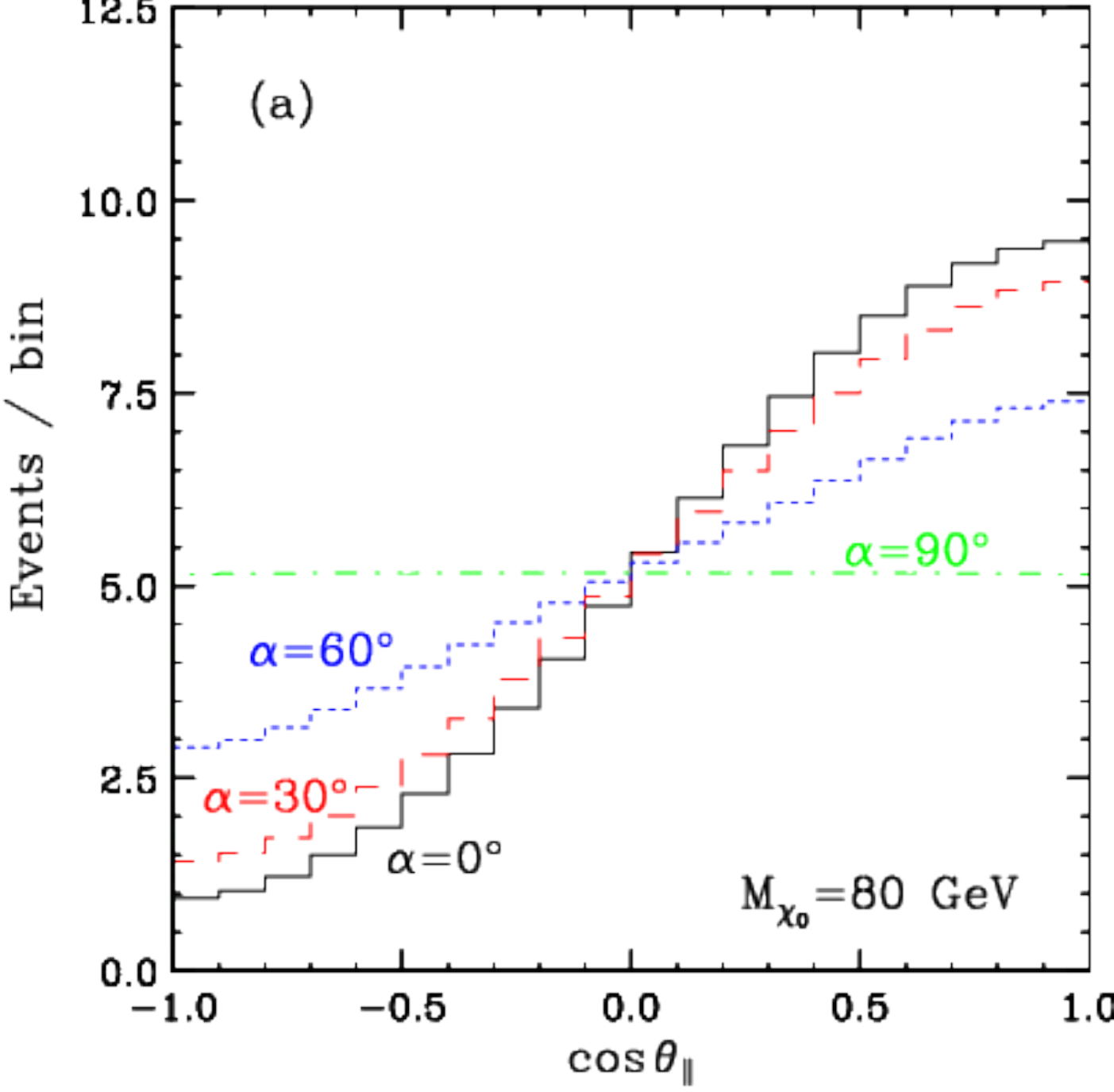} \hspace{0.5cm}
\includegraphics[scale=0.5]{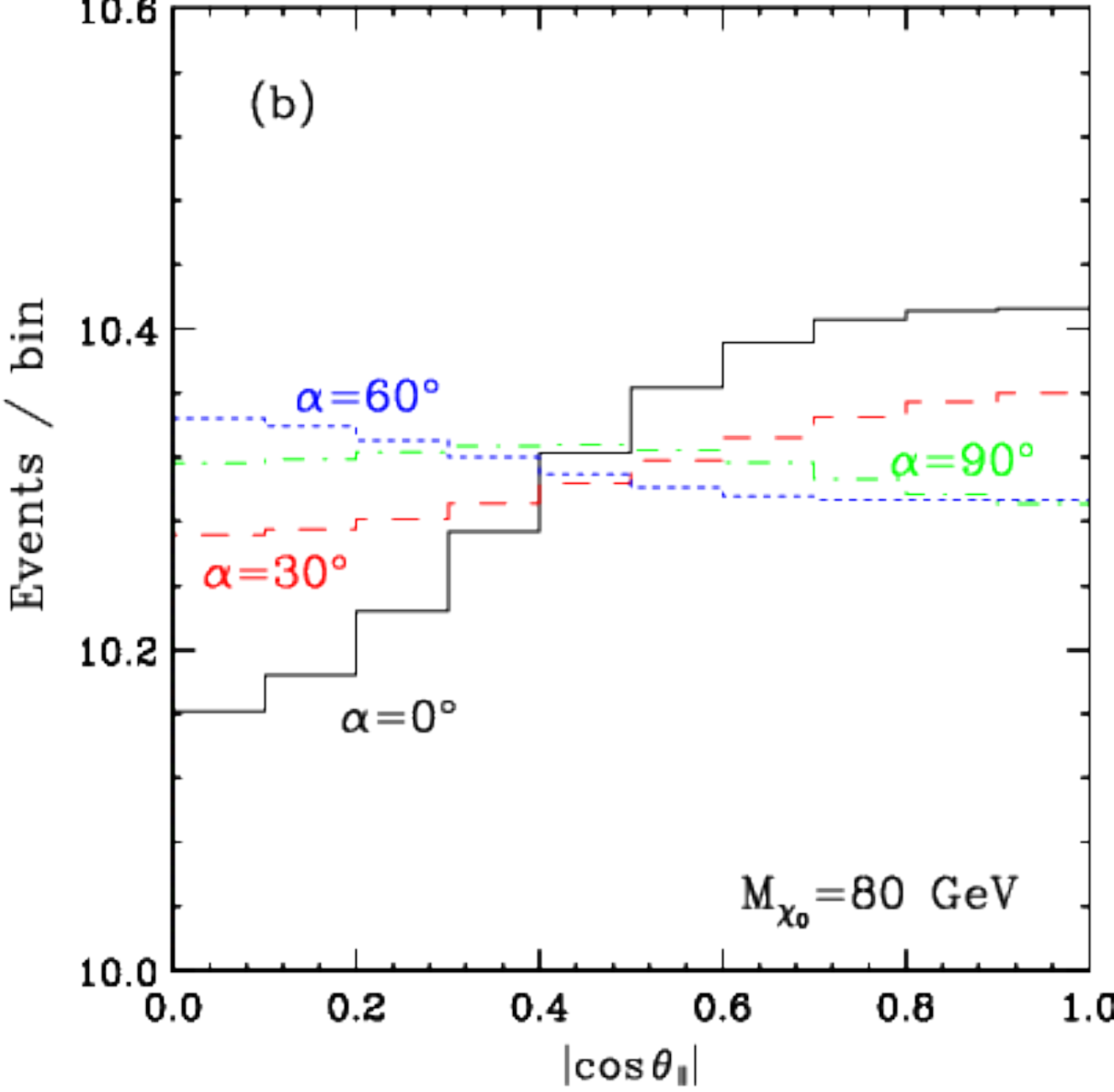}}
\caption{Angular distribution of nuclear recoil events with (a) and without (b) head-to-tail capability, rotated at different angles with respect to the incoming WIMP trajectory for a 80 GeV WIMP. $\alpha=0^\circ$ and $\alpha=90^\circ$ have been discussed earlier. }
\label{cosperpalpha}
\end{figure}

Figure \ref{corrhtnht} illustrates the double differential distributions for the first two types of detector concepts (head-to-tail or no head-to-tail) for both a parallel and perpendicular electric field. 
We define them as 
$  \frac{d^2 N}{ d E_R d \cos \theta_\parallel      } $ in (a),
$  \frac{d^2 N}{ d E_R d | \cos \theta_\parallel | } $ in (b),
$ \frac{d^2 N}{ d E_R d \cos \theta_\perp    }$ in (c) and 
$  \frac{d^2 N}{ d E_R d | \cos \theta_\perp |} $ in (d), respectively, 
where subscripts $\parallel$ and $\perp$ denote the direction of the drift electric field with respect to the WIMP direction. 
Along each curve, the same number of events are expected. Figure \ref{corrhtnht}(a) is the most ideal case with a full coverage of the recoil angle.
By our set up, 
$  \frac{d N}{ d \cos \theta_\parallel     }   =  \frac{d N}{ d \cos \theta     } $ and also 
$  \frac{d N}{ d | \cos \theta_\parallel  |  } = \frac{d N}{ d | \cos \theta |    } $, which is 
the `folded' directional recoil rate, where $|\cos \theta|$ does not distinguish the beginning of the recoil track from its end (lack of head-tail discrimination) \cite{Alenazi:2007sy, Billard:2012bk}. 
A detector that is fixed on Earth may weaken the DM directionality and we have investigated this effect in our simulation by orienting the electric field at a fixed angle $\alpha$ with respect to the incoming WIMP direction as shown in Figure \ref{cosperpalpha}. The $\alpha=0$ case corresponds to a movable detector that we have described and the detector that is fixed on Earth would include a combination of different $\alpha$ angles, washing out the angular information. As the movable system provides the best sensitivity, we will consider this case throughout the paper.

Note that our study point $5\times 10^{-11}$ pb  for a light dark matter particle falls within the overwhelming neutrino backgrounds in direct detection experiments as described in \cite{Billard:2013qya}. The effects of neutrino backgrounds on directional detection have been partially studied in Ref. \cite{Grothaus:2014hja}. 

To determine the dark matter mass, cross section, and anisotropy, we perform simulations for these types of detectors. We assume an energy threshold of $4$\,keVnr (unless noted differently). Gaussian smearing is applied for both energy and angle as follows: 
\begin{equation}
F(E,\theta) = \int F(E',\theta') 
\left ( \frac{1}{ \sigma_{E}\sqrt{2\pi} } \, e^{-\frac{  ( E - E'  )^{2}}{2\sigma^2_{E}} } \right )
\left ( \frac{1}{ \sigma_{\theta}\sqrt{2\pi} } \, e^{-  \frac{ ( \theta - \theta' )^2 }{2\sigma^2_{\theta}} } \right )
dE'  d\theta' \, , \label{res} 
\end{equation}
where $F(E,\theta)$ is event rate function (Eq. \ref{doubleratereq}), $\sigma_{E} = \lambda \sqrt{E}$ is the energy resolution and $\sigma_{\theta}$ is a constant angular resolution. 
%
%
We have assumed $\lambda=1$ and $\sigma_\theta = 30^\circ$ in our numerical study, unless noted otherwise.
They are rather conservative choice compared to those reported in literature (see Ref. \cite{Ahlen:2009ev} for details.).
In the case of low energy recoils we would have to worry about negative energies in the above Kernel, but we found that a threshold cut at 4 keVnr is large enough to avoid such events. The angular smearing was carried out in $\theta'$-space using the Kernel in Eq. \ref{res}. For a given number of events at an angle $\theta$-bin ($0 < \theta < \pi$), the smearing Kernel is applied to a large array of linear angles-bins of $\theta'$. The events that fall below 0 and above $\pi$ respectively are then folded back on the main range of the distribution. This is done to preserve the angular range of the original $\theta$ distribution and in this way the total number of events is conserved as required. We choose a cross-section of $5\times 10^{-11}$ pb for simulation purposes (unless noted otherwise), which roughly gives 103 (143) events after (before) the 4 keV threshold cut in a Xenon detector for 10 ton-year, assuming zero background.

\subsection{Estimation of WIMP Mass and WIMP-Nucleon Cross-Section\label{sec:param}}
\begin{figure}[ht!]
\centering
\centerline{ \includegraphics[scale=0.46]{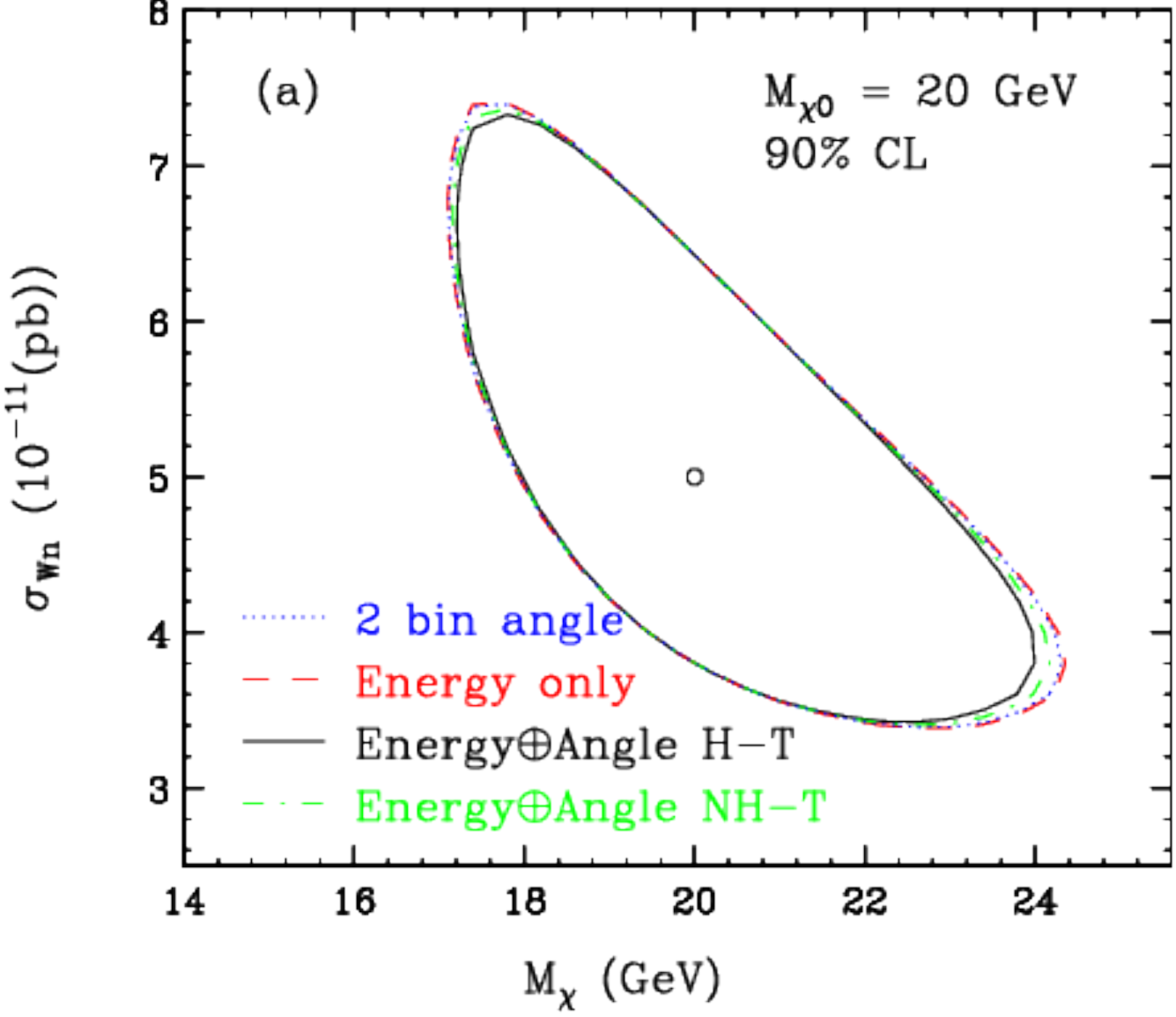} \hspace{0.35cm}
                   \includegraphics[scale=0.46]{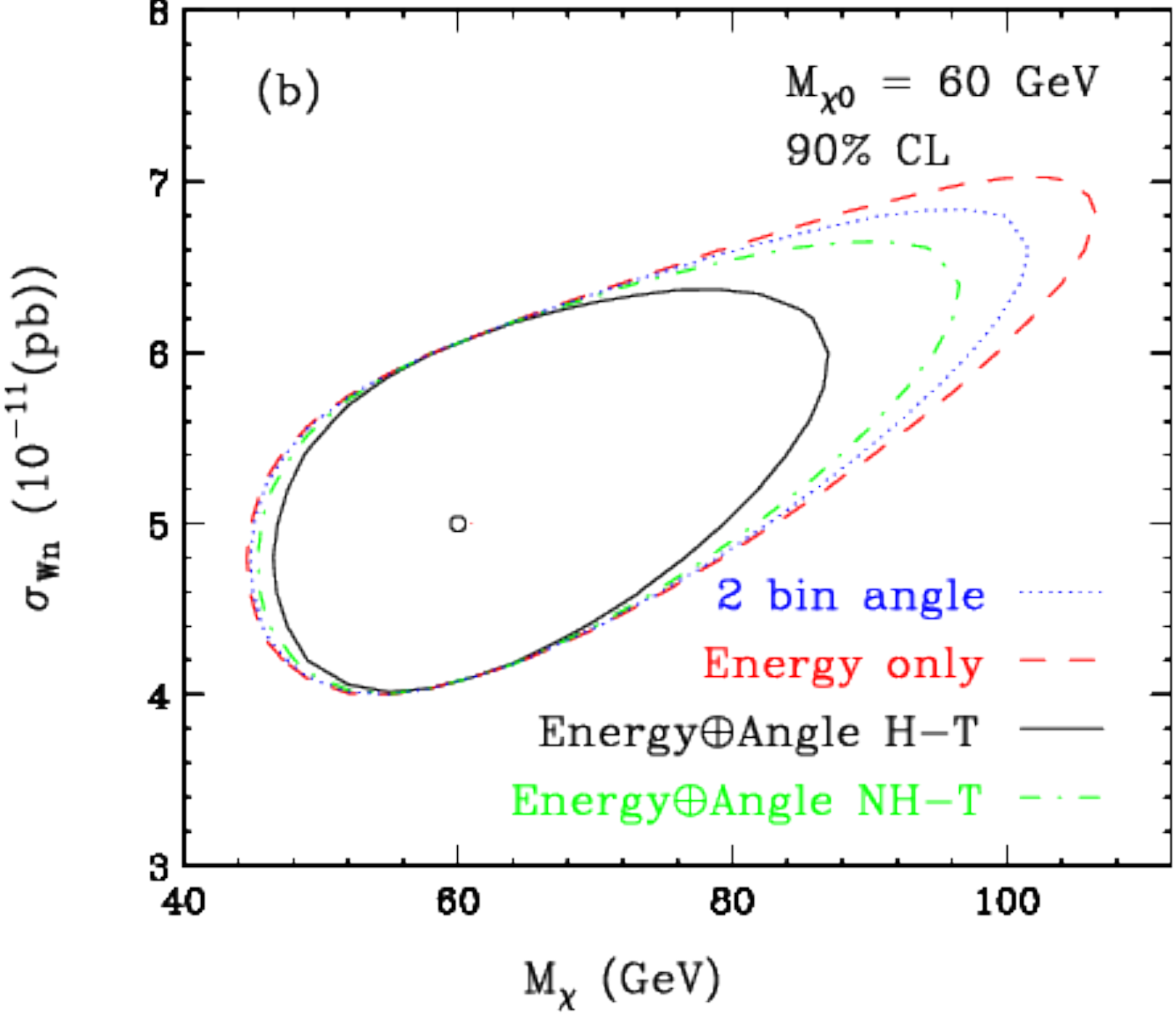}   }
                   \vspace{0.5cm}
\centerline{ \includegraphics[scale=0.46]{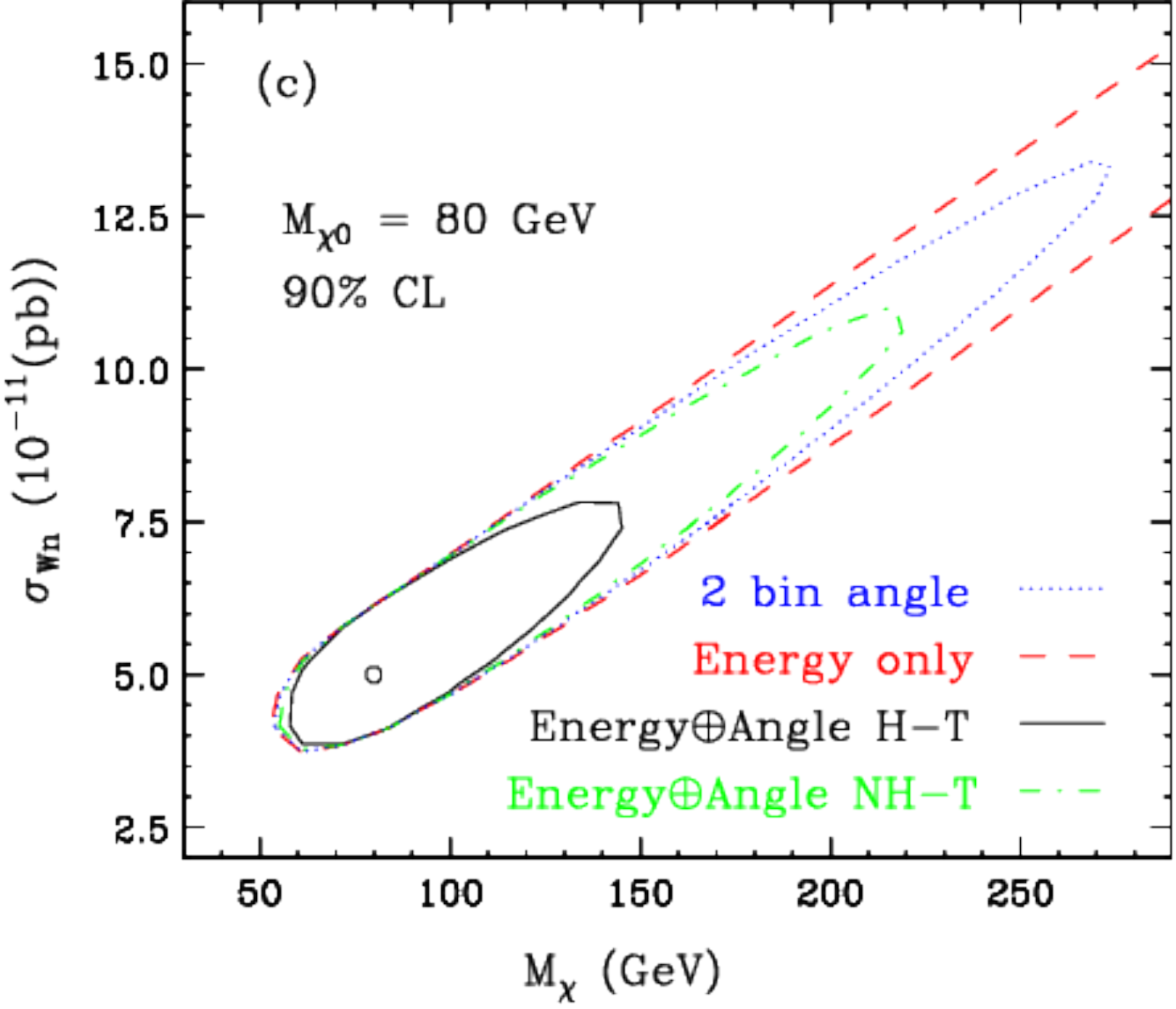} \hspace{0.35cm}
	            \includegraphics[scale=0.46]{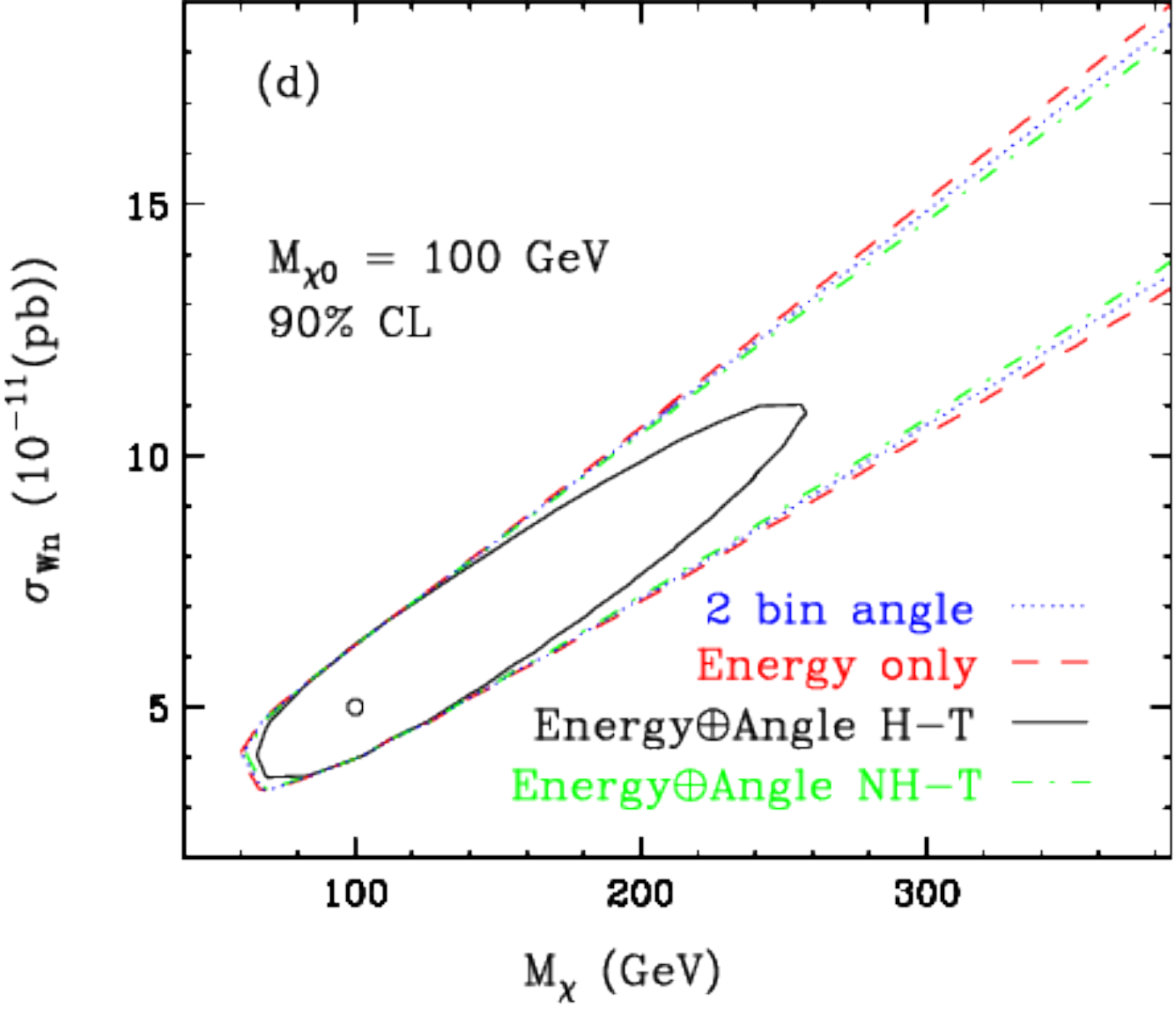} }
\caption{
Illustration of measurements of mass and cross section of galactic dark matter at 90$\%$ confidence level (CL) with $\sigma_{W n 0} = 5 \times 10^{-11}$ pb for (a) $M_{\chi_0}=15$ GeV, (b) $M_{\chi_0}=60$ GeV, (c) $M_{\chi_0}=80$ GeV, and (d) $M_{\chi_0}=100$ GeV. The four different contours represent different types of detectors assumed in the likelihood analysis. 
For `Energy$\oplus$Angle' (black, solid), we use both recoil energy and angular information obtained from the theoretical distribution. 
We integrate over the angle and annual modulation to obtain the recoil energy information only for `Energy only' (red, dashed). 
A detector without head-to-tail information is shown in the green-dot-dashed contours. 
Finally for `2 Angle Bins' (blue, dotted) we use 2 bins in the angular distribution. The input point for our simulation is ($M_{\chi 0}$, $\sigma_{W n 0}$) and is represented by a dot inside the ellipses. All events are normalized to 10 ton-year exposure for a Xenon detector, 
including detector resolution effects and 4 keV threshold cut.}
\label{CLcontours}
\end{figure}

To illustrate the capability of measurement of the WIMP mass and cross section for the detectors discussed in the previous section, we calculated expected event rates using Eq. (\ref{doubleratereq}) for four different study points 
$M_{\chi_0}=20$ GeV, $M_{\chi_0}=60$ GeV, $M_{\chi_0}=80$ GeV and $M_{\chi_0}=100$ GeV for a fixed input cross section of $\sigma_{W n 0} = 5 \times 10^{-11}$ pb, as shown in Figure \ref{CLcontours}. The event rates are normalized to a 10 ton-year exposure of the Xenon detector. 
%
%
%
The physics information (dark matter mass and interaction cross section) is determined using a binned likelihood analysis assuming a Poisson probability distribution of our signal events:
\begin{equation}
L = \prod_{i=1}^{N_{bin}} \frac{ ( N_{E}^i )^{N_{O}^{i}}}{ (N_{O}^{i})! } \exp^{-N^{i}_{E}} \, ,
\end{equation}
where $N^{i}_{E}$ is the expected number of (template) events, $N_{E} = N_{E} (M_{\chi}, \sigma_{W n}) $ and $N_{O}^{i}$ is the observed number of (signal) events in each bin. 

Four different detectors are examined:
(i) a conventional non directional detector where only the recoil energy of the events is measured (denoted as `Energy only') shown as red-dashed ellipses in Figure \ref{CLcontours},
(ii) a detector which has the ability to measure the energy, angle and annual modulation signal of every event (denoted as `Energy$\oplus$Angle') 
shown as the black-solid ellipses, 
(iii) a detector without head-to-tail information (shown in green-dot-dashed ellipses), and 
(iv) a detector in which we do not have the ability to measure the angle of each event, but we can determine the number of events within a certain angular `cone', i.e. we split the angular distribution of events in two bins, and use both bins in the likelihood analysis, but since we cannot determine the precise angle of each and every event, we only know that an event fell in this angular space of a certain size (with $\pm30^\circ$ opening angle for this study) or outside (2 bin angle system). 
The `2 Angle Bins' case in blue-dotted ellipses represent the performance of this type of detector.  
For representation purposes, we do not mention annual modulation in the figures, 
since the results do not change much whether the annual modulation effect is included or not.
%
We carry out the binned likelihood analysis and 
obtain a region of the parameter space that is consistent with the input point at 90\% C.L, shown as four ellipses for each case in Figure \ref{CLcontours}. The minimum of the log-likelihood is marked for each case and they should coincide with the input study point in the absence of any statistical fluctuations.
Although finite statistics would shift the best fit point off from the original input and may alter the shape of contours slightly, 
our study indicates what improvement is expected in the best case scenario.

The difference in the orientation of the ellipses for the 20 GeV case and the rest is easily understood from the interplay between a threshold cut and the shape of the differential energy rate. For a light DM of mass $M_{\chi_0}$ (20 GeV in this case), the differential distribution of recoil energy is very steep and a majority of the events are cut away with a threshold cut, which implies that 
one needs a higher cross section to fit the data with $M_\chi < M_{\chi_0}$. 
On the other hand, 
the fitting procedure requires a smaller cross section, as the energy distribution is less steep for $M_\chi > M_{\chi_0}$.
This is shown in Figure \ref{CLcontours}(a). However this is no longer true, if the input mass $M_{\chi_0}$ is large as illustrated in Figure \ref{CLcontours}(b)-(d).

\begin{figure}[t]
\centering
\includegraphics[width=7cm,height=6.5cm]{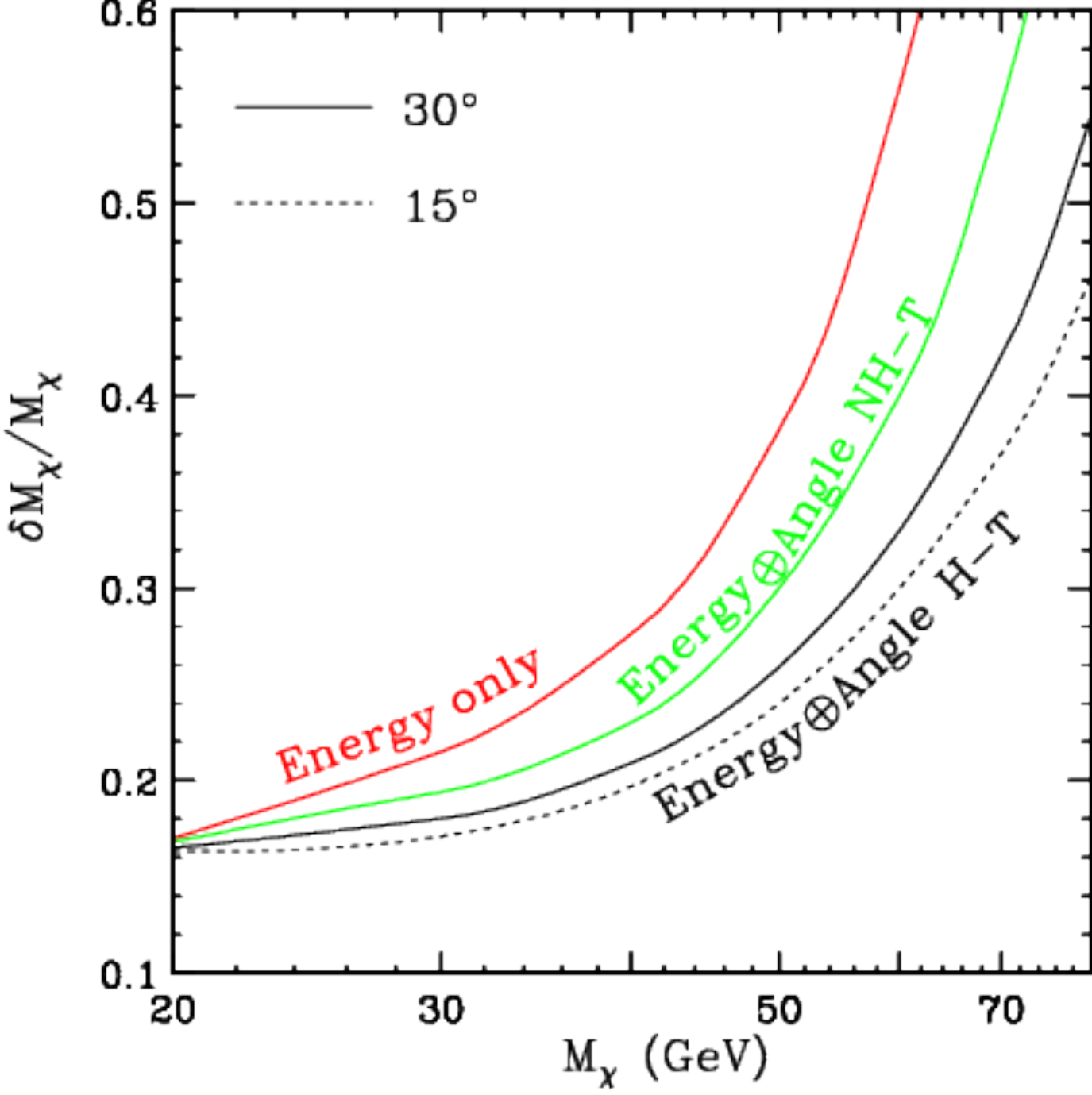}
\caption{Relative WIMP mass uncertainty as a function of WIMP mass. 
We compare three detectors, 
the conventional (non-directional) detector (in red),  one with full coverage of directionality (in black), and the other without head-to-tail information (in green).
The case with 2 angle bins lies between the conventional case and the no head-to-tail case and is not shown here.
A $15^{\circ}$ ($30^{\circ}$) angular resolution is assumed for the dotted curve (solid, both green and black).}
\label{mass_err}
\end{figure}

We also notice that for a light DM, directional information does not play an important role in measurements of parameters.  
However for a heavy dark matter (heavier than 100 GeV), the full directionality is crucial in precision measurements.
Precision can be substantially improved for the intermediate mass range below 100 GeV even without head-tail information.
We illustrate this in Figure \ref{mass_err} where the relative WIMP mass uncertainty ($\delta M_\chi / M_\chi$) is shown as a function of WIMP mass for different classes of detectors.
%

%

We have also studied an impact of both angular resolution and energy resolution to see the effect of finite resolution on the dark matter parameter determination and we find that the 90$\%$ contours do not change significantly. Also in the case of a detector with the 2 bin angular system described above, we used a benchmark angular opening of $30^{\circ}$ in Figure \ref{CLcontours}. We then tested for different sizes of angular area, i.e.  $60^{\circ}$ and $90^{\circ}$, but we found no large difference in neither the angular distributions nor the 90$\%$ CL contours. Our results imply that directionality with full angular coverage improves the measurement of masses and cross section significantly especially for a heavy dark matter. 
A detector without head-to-tail information or one with limited 2 angular bins provides a marginal improvement in the accuracy of the parameter determination.

\subsection{Angular Distributions and Anisotropy of Dark Matter Flow  \label{sec:anglular}}

A non-trivial angular dependence of nuclear recoils produced in dark matter interactions arises due to an asymmetric velocity distribution of dark matter in the lab-frame. As mentioned in the previous section, we use the truncated Maxwell-Boltzmann distribution in Eq. (\ref{maxwell}) as our reference. Therefore in this set up, the asymmetry shows up entirely due to the motion of the detector as shown in Eq. (\ref{doubleratereq}), i.e., 
$\frac{d R}{d \cos \theta }$ is isotropic (flat) for $v_E = 0$. 

\begin{figure}[t]
\centerline{
\includegraphics[scale=0.48]{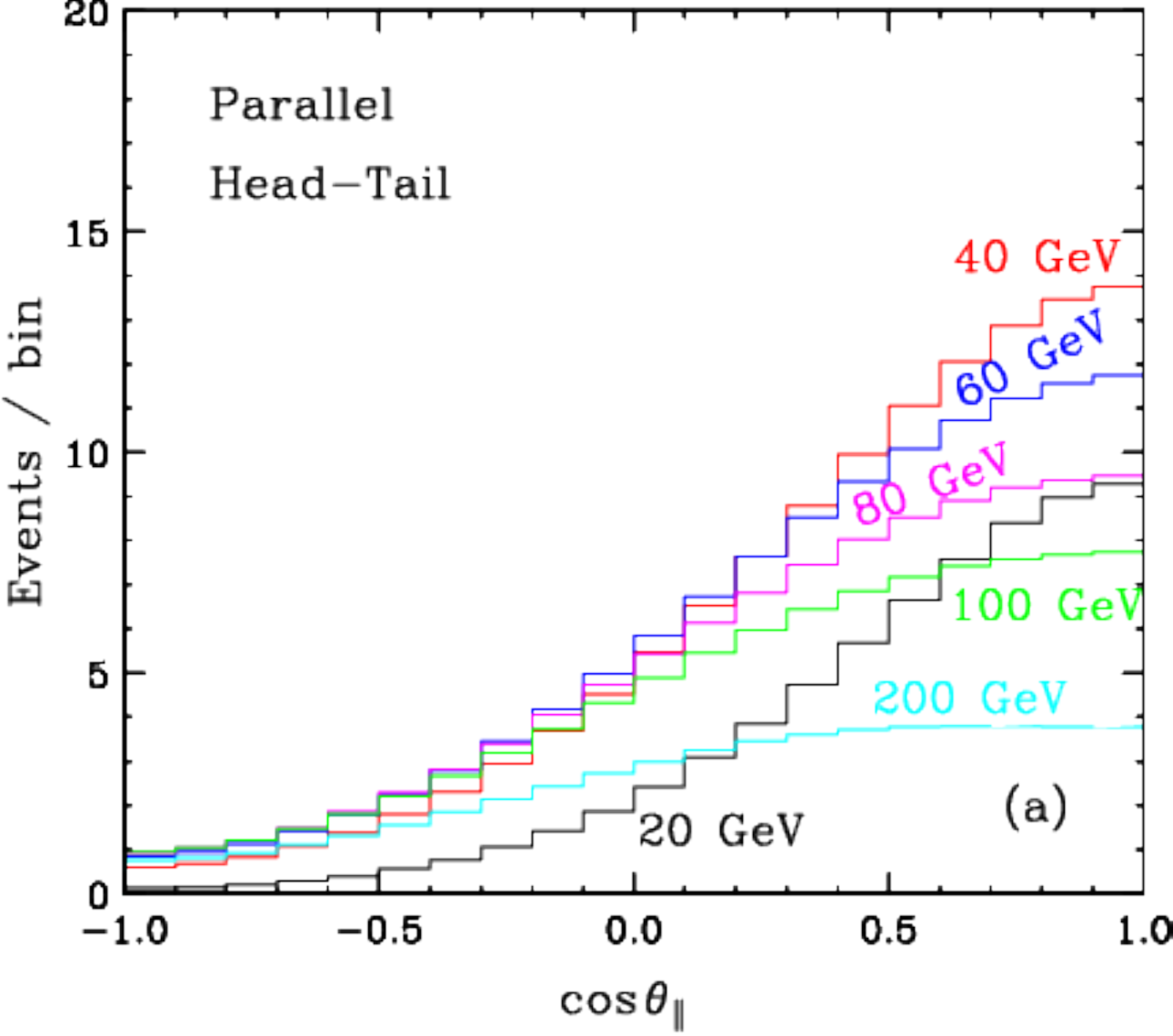} \hspace{0.3cm}
\includegraphics[scale=0.48]{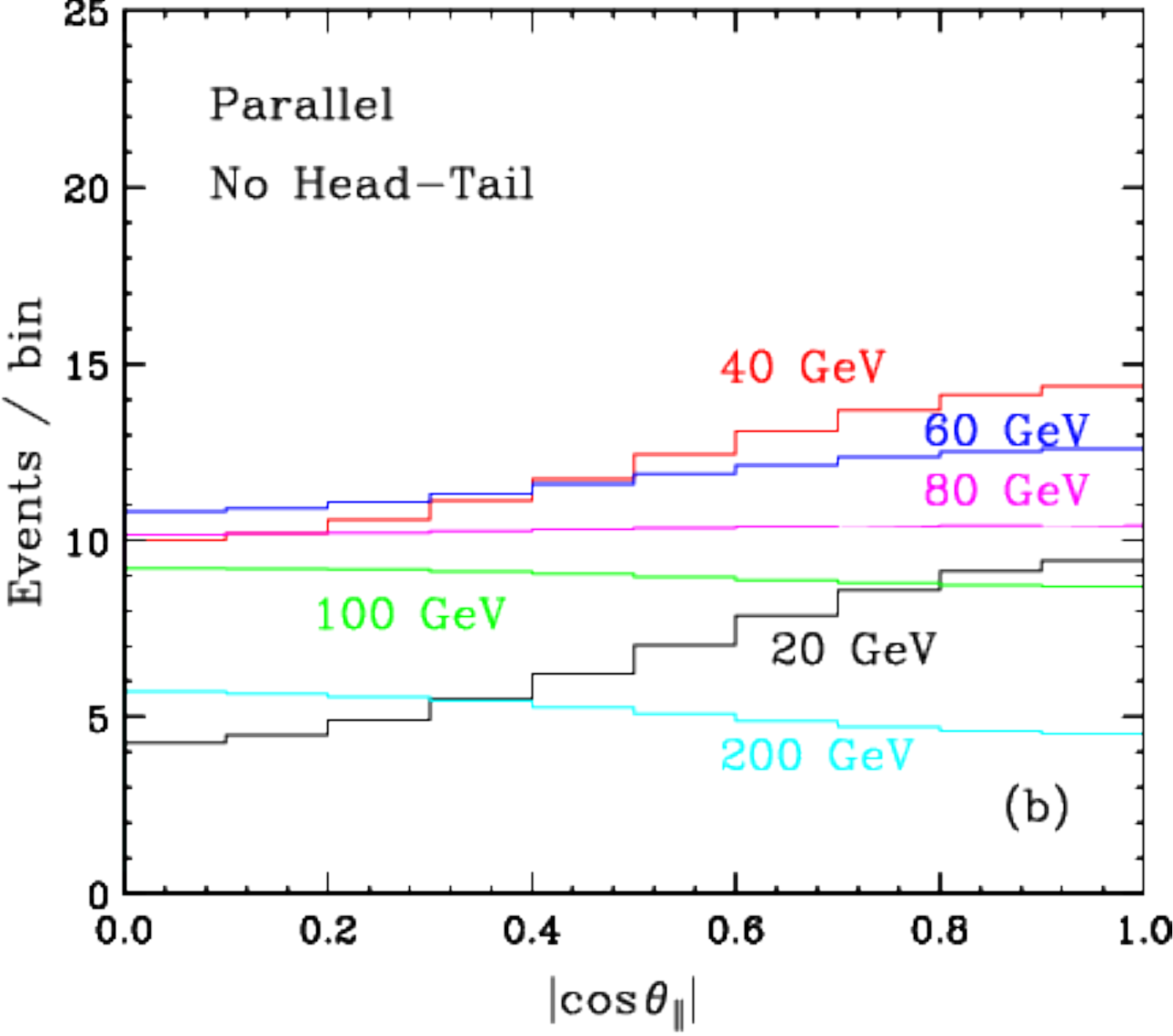}}
\vspace{0.5cm}
\centerline{
\includegraphics[scale=0.48]{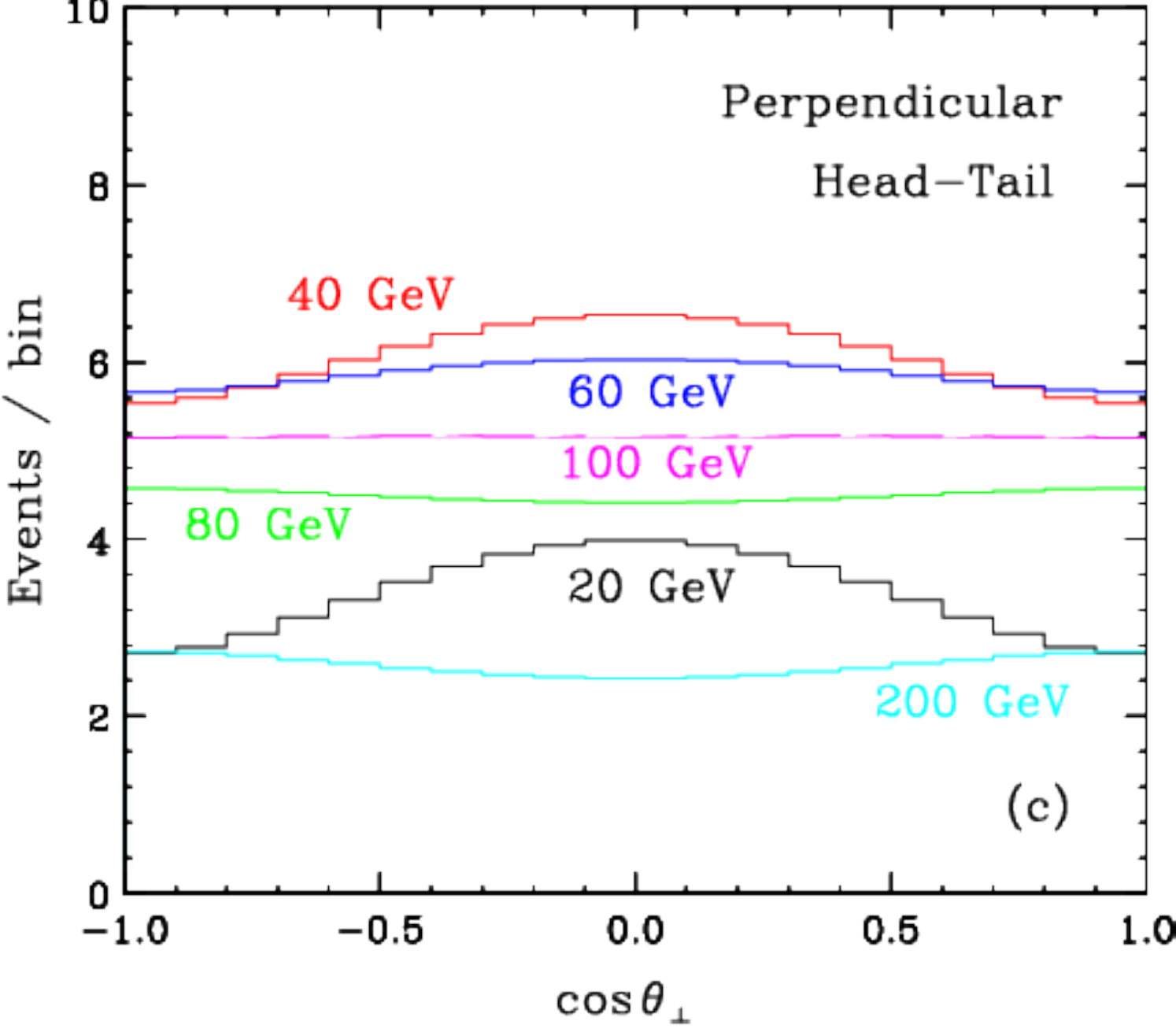} \hspace{0.3cm}
\includegraphics[scale=0.48]{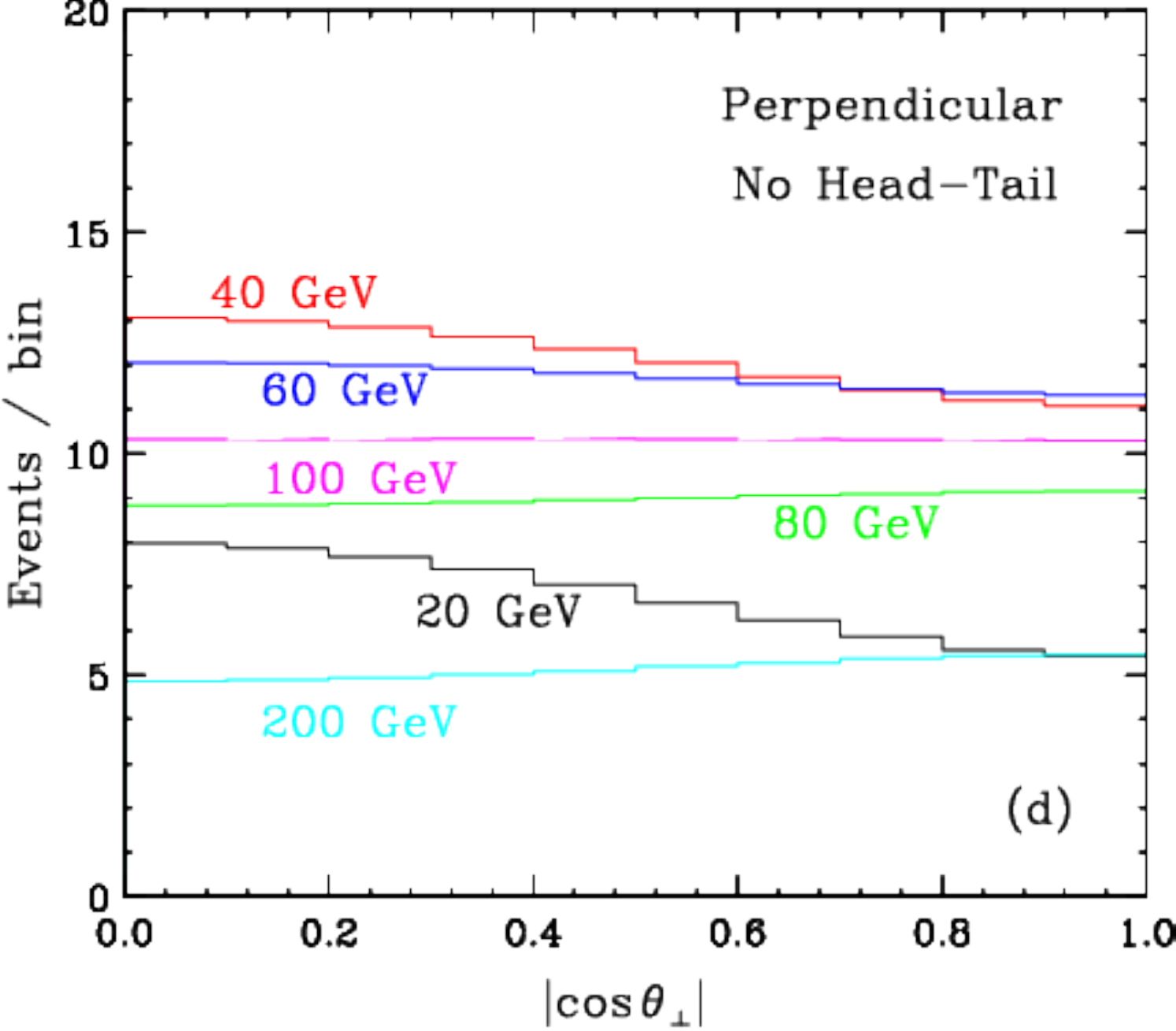}}
\caption{Comparison between a detector with capabilities of measuring head-to-tail (left) and one which is incapable of such a distinction (right), for a 4 keV energy threshold cut, assuming the Drift Electric field is pointing parallel (top) or perpendicular (bottom) to the initial WIMP trajectory. 
}
\label{parapara}
\end{figure}

To maximize the observed anisotropy we constantly adjust the orientation of our detector with respect to the Cygnus direction. 
As an exercise we have studied two detector configurations, one with the drift field parallel to the dark matter direction and one with the drift field perpendicular to it.
In Figure \ref{parapara}, we show angular distributions of two types of detectors based on the capability of head-to-tail discrimination for several choices of dark matter mass. The two plots in the top panel correspond to 
$ \frac{d R}{d \cos \theta_\parallel } $ in (a) and $\frac{d R}{d | \cos \theta_\parallel |}$ in (b), 
which is $ \frac{d R}{d | \cos \theta |}$, also known as the `folded' differential in Ref. \cite{Alenazi:2007sy}.
%
%
A precise comparison would be somewhat difficult since we assume a different set up and different materials than those used in Ref. \cite{Alenazi:2007sy} (CS2 and CF4).
However we are able to reproduce a (roughly) consistent result and especially the shape of our folded distribution resembles that in Ref. \cite{Alenazi:2007sy}.
In our set-up, this folded distribution is obtained when the drift electric field is parallel to the initial WIMP direction based on details of the columnar recombination effect. 
$  \frac{d R}{d \cos \theta_\perp } $ and $ \frac{d R}{d | \cos \theta_\perp |} $ are shown in (c) and (d), respectively.

\begin{figure}[t]
\centerline{
\includegraphics[scale=0.32]{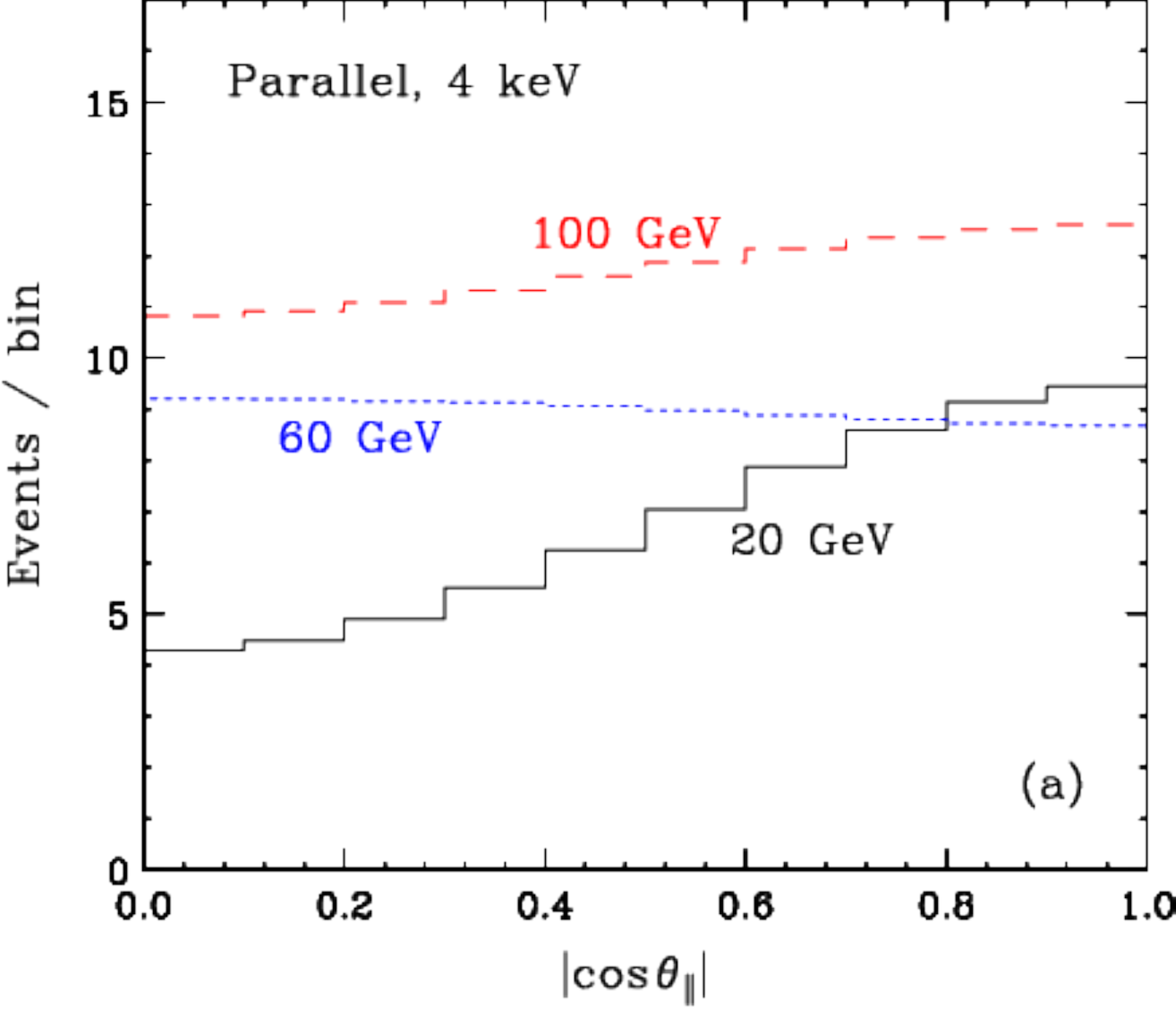} \hspace{0.04cm}
\includegraphics[scale=0.32]{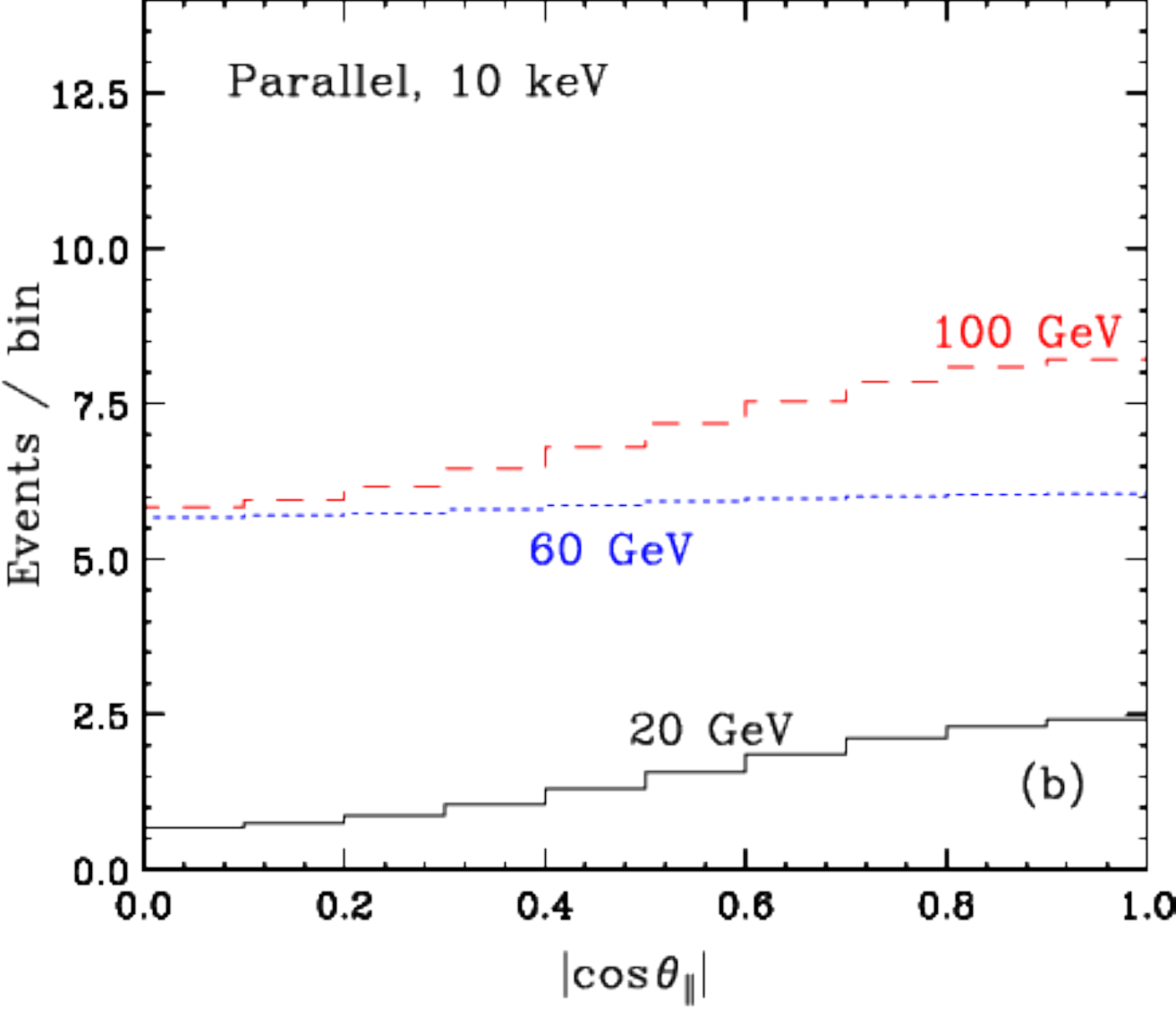}\hspace{0.04cm}
\includegraphics[scale=0.32]{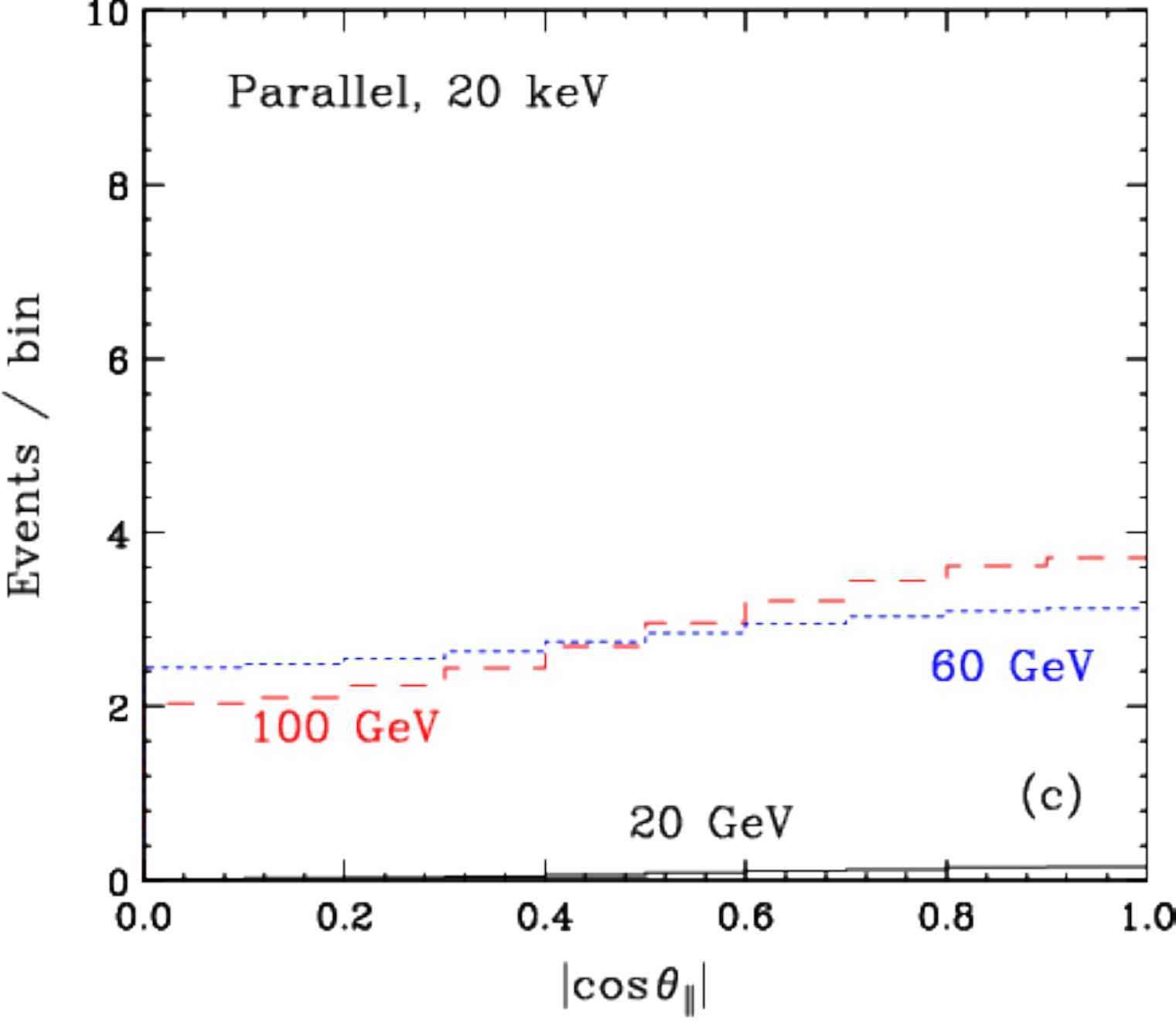} } 
\vspace{0.3cm}
\centerline{
\includegraphics[scale=0.32]{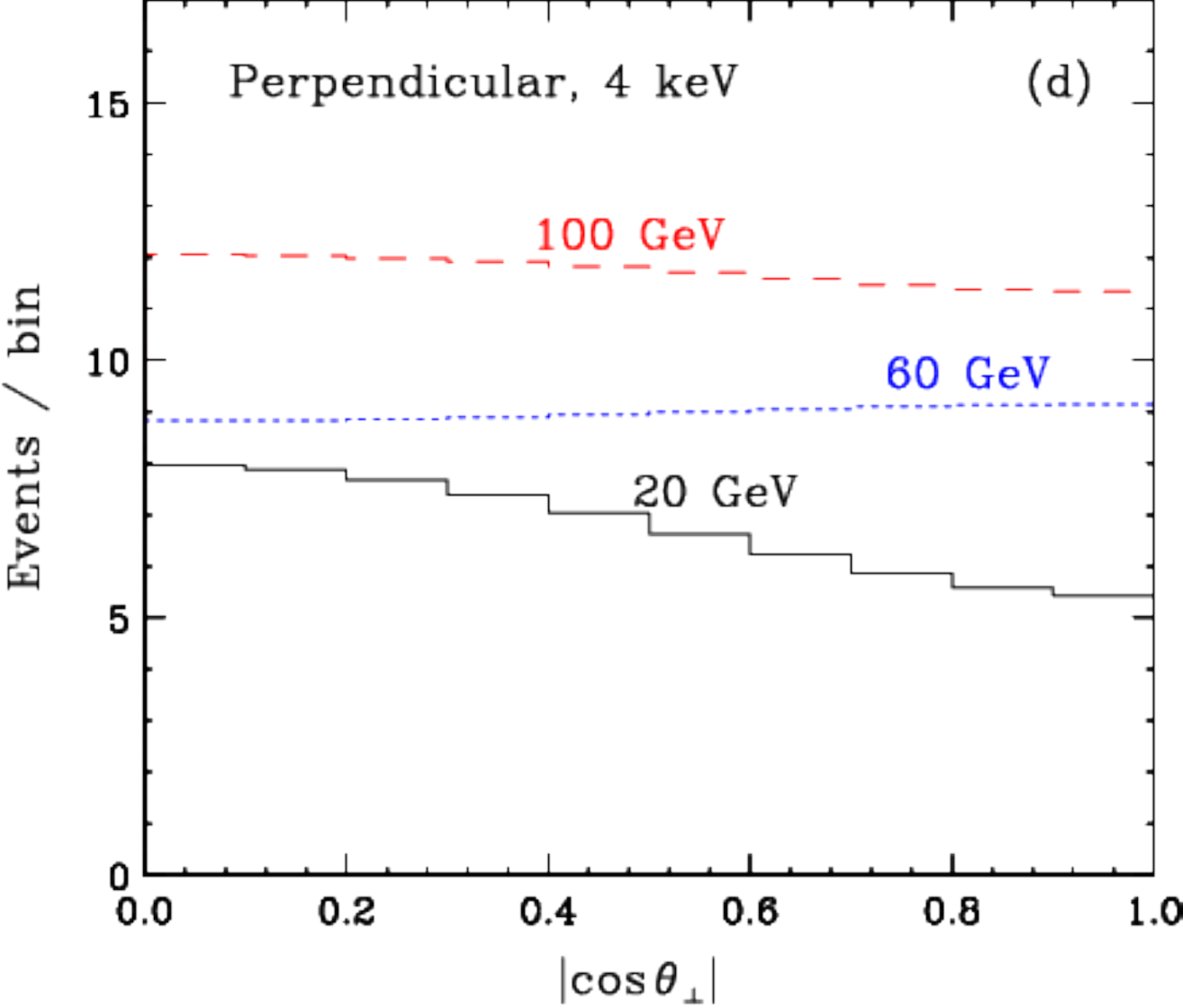} \hspace{0.03cm}
\includegraphics[scale=0.32]{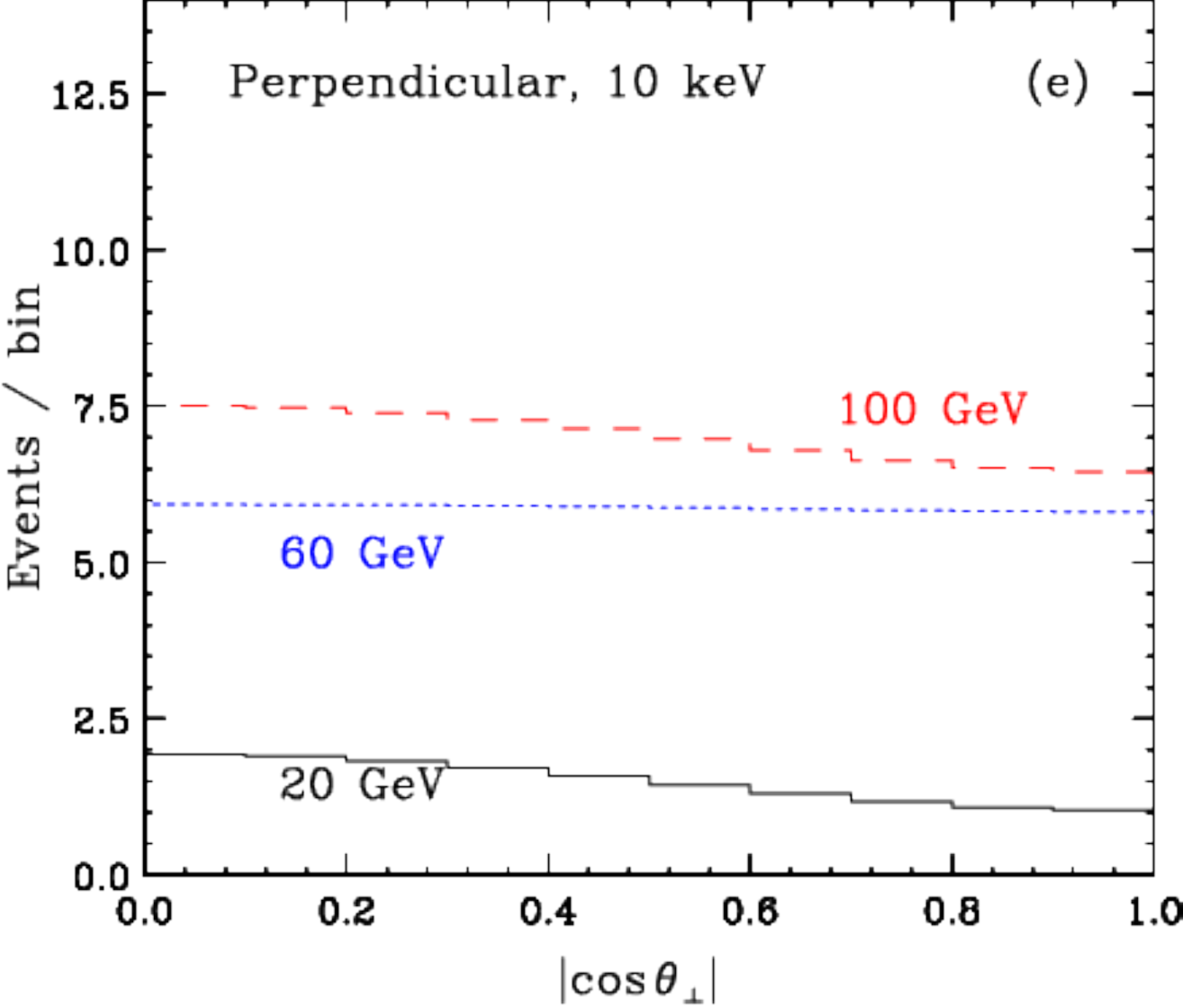} \hspace{0.03cm} 
\includegraphics[scale=0.32]{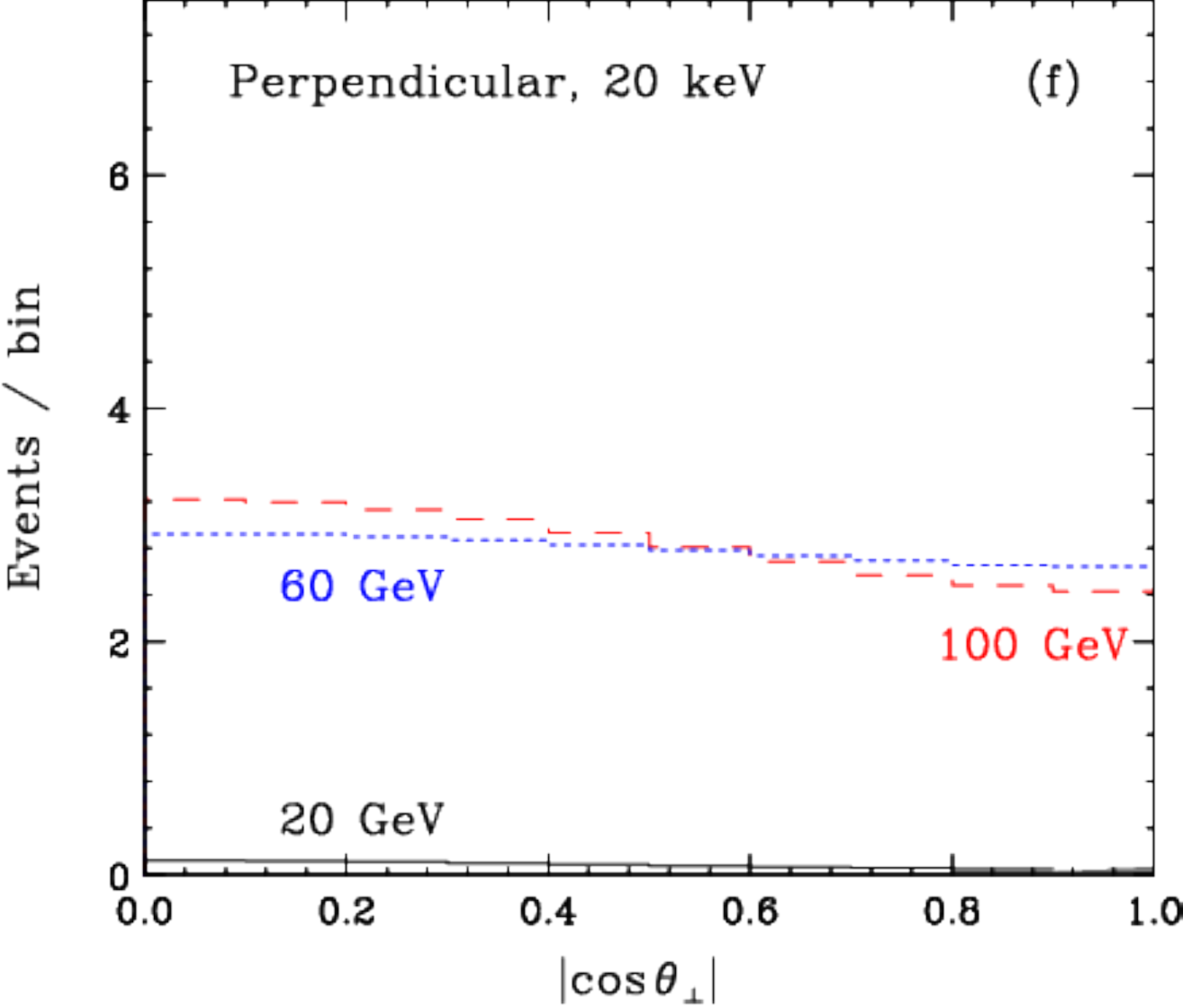}}
\caption{Angular distributions for a drift electric field oriented parallel (top) and perpendicular (bottom) to the initial WIMP trajectory for a detector with no head-to-tail capabilities. These are illustrated for 4 keV, 10 keV and 20 keV energy threshold cuts for 20 GeV, 60 GeV and 100 GeV WIMP masses.}
\label{para41020}
\end{figure}

As shown in Figure \ref{parapara}(a), dark matter scatters predominantly in the forward direction and details of the shapes of the angular distributions are dependent on the mass of dark matter and the imposed threshold cut. Unfortunately a lack of head-tail discrimination puts a severe hurdle on the measurement of the anisotropy of the dark matter flow (see (b) and (d)).
The situation may be (slightly) improved for a light dark matter (20 GeV and 60 GeV) at the cost of signal statistics by imposing a higher threshold cut as shown in Figure \ref{para41020}, while there is no change for a relatively heavy dark matter (100 GeV). This is due to a correlation between the recoil energy and the scattering angle. 
Angular distributions observed in a detector with the field perpendicular to the average WIMP direction also show a similar behavior but the sensitivity of the measurement would be greatly reduced.

%

\subsection{Anisotropy with a Columnar Recombination Detector\label{sec:cr}}

Our discussion in the previous section is somewhat generic in a sense that results do not particularly utilize the effect of CR. In a real experiment, the CR detector would not measure recoil energy and angle directly, but would rather count the number of electrons and photons released from Ionization ($I$) and Scintillation ($S$) processes respectively, with some detector resolution and efficiency. 
The recoil energy and the recoil angle are obtained as a function of the two variables $I$ and $S$. The efficiency of the measurement of $I$ and $S$ will also depend on the orientation of the drift $ \vec {E}$ with respect to the nuclear recoil trajectory. If the field is parallel to the recoil trajectory, one expects a higher rate of ionization and scintillation, and the opposite effect when the field is perpendicular to the recoil trajectory. After these are measured one can convert these observables to recoil energy and recoil angle. We assume the following relation.
\begin{eqnarray}
S &=& F(E_{R}) \,  \epsilon \, E_{R} \cos^{2} \theta_{L} \, , \\ 
I &=& F(E_{R}) \, \epsilon \,  E_{R} \sin^{2} \theta_{L}  \, ,
\end{eqnarray}
where $F(E_{R})$ is the number of observed photo-electrons per keV which takes into account the quenching factor, and $\epsilon$ is the detection efficiency of photons. For the $F(E_{R})$, we have adapted experimental values for absolute S1 (prompt scintillation) yields for electron recoils in Xenon as in Ref. \cite{Szydagis:2011tk}. Absolute yield (in photons/keV) is given as a function of the incident gamma energy compared with their Monte Carlo output taking into account the recombination probability. We have taken the best reproduction based on their model and scaled it down with our choice of $\epsilon$. Smearing $S$ and $I$ with Poisson statistics is performed before converting back to energy and angle.

\begin{figure}[t]
\centerline{
\includegraphics[scale=0.48]{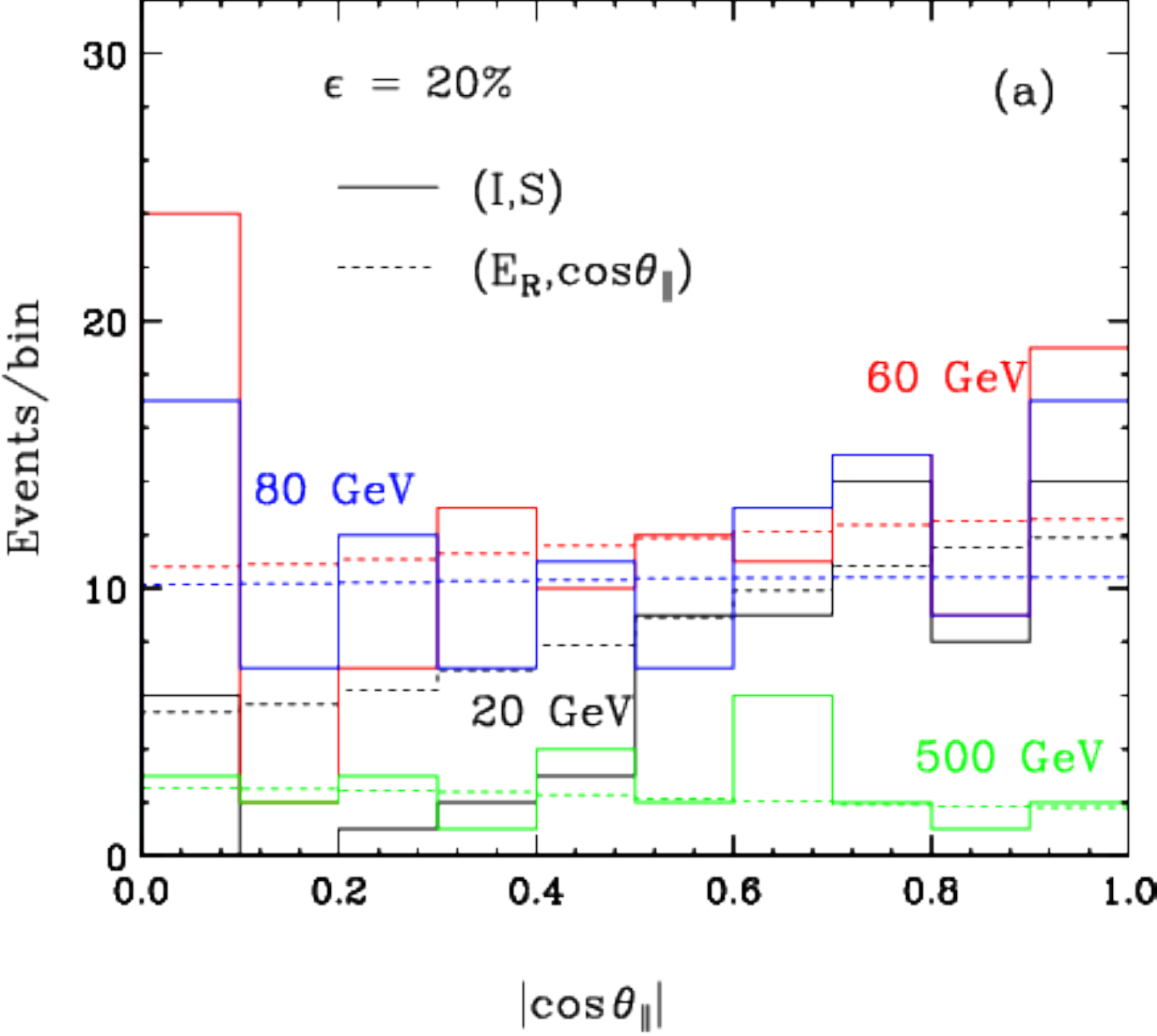}\hspace{0.4cm}
\includegraphics[scale=0.48]{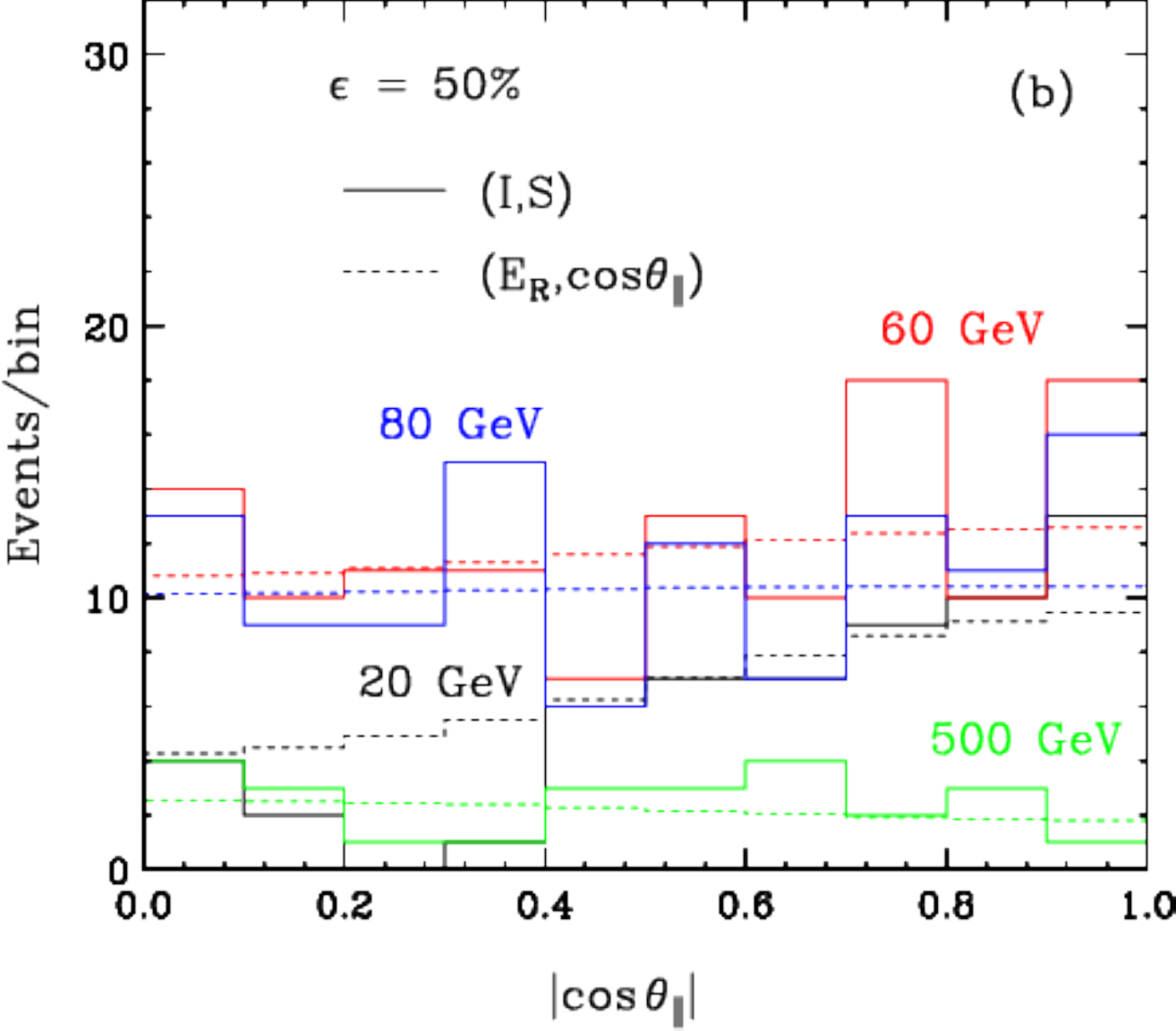}
}
\caption{Angular distribution, $ \frac{ d N}{ d |\cos\theta_\parallel |} $, for a realistic CR detector for two different photo detection efficiencies $\epsilon$. The solid histograms are angular distributions obtained from the ionization and scintillation light assuming a detector resolution with Poisson distribution using, $\cos\theta_{L} = \sqrt{S/(S+I)}$. The dashed histograms are events generated from a Monte-Carlo simulation based on the theoretical distribution, assuming a detector resolution in Eq \ref{res}.}
\label{cosSI}
\end{figure}

\begin{figure}[t]
\centerline{
\includegraphics[scale=0.33]{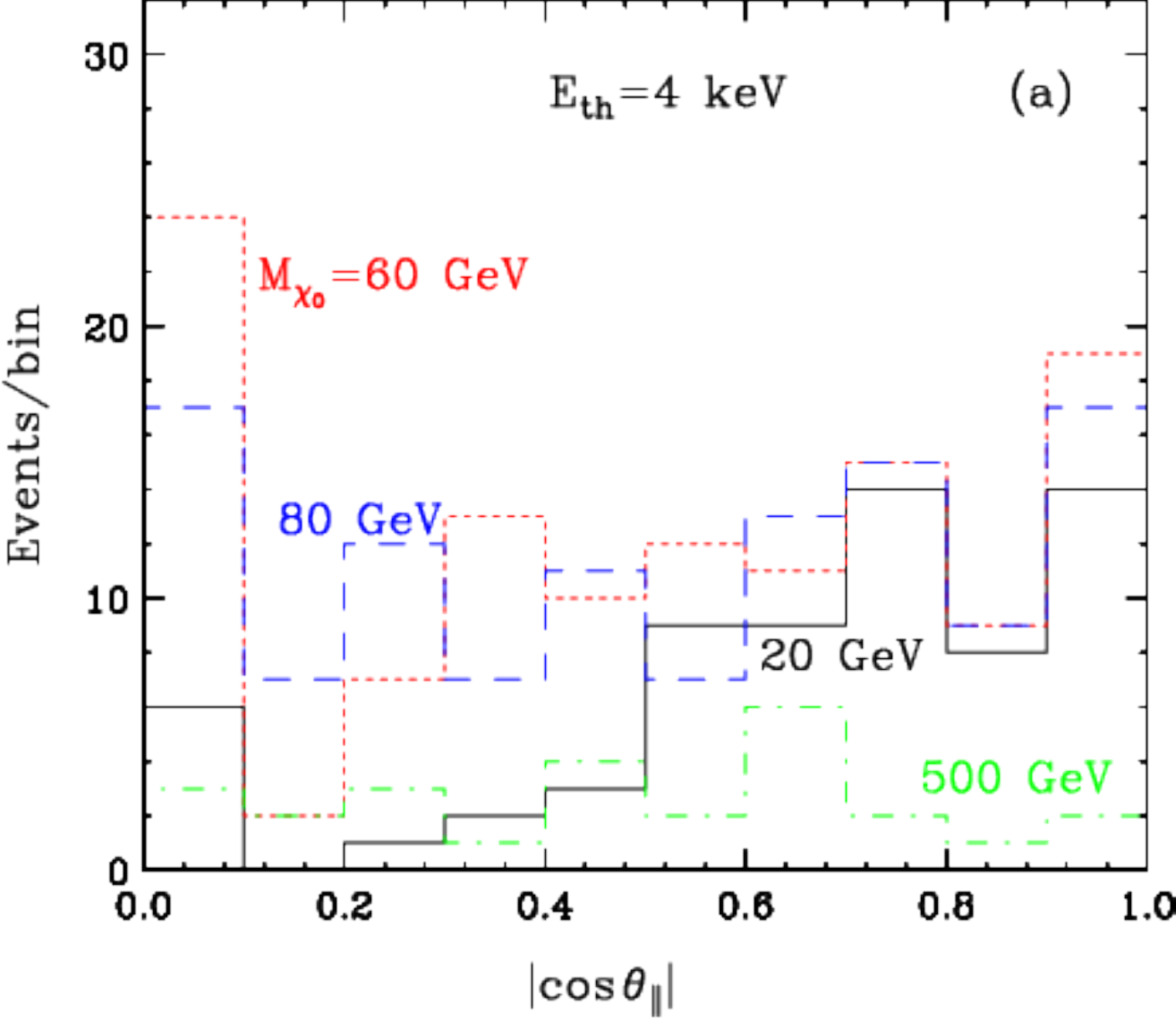}\hspace{0.07cm}
\includegraphics[scale=0.33]{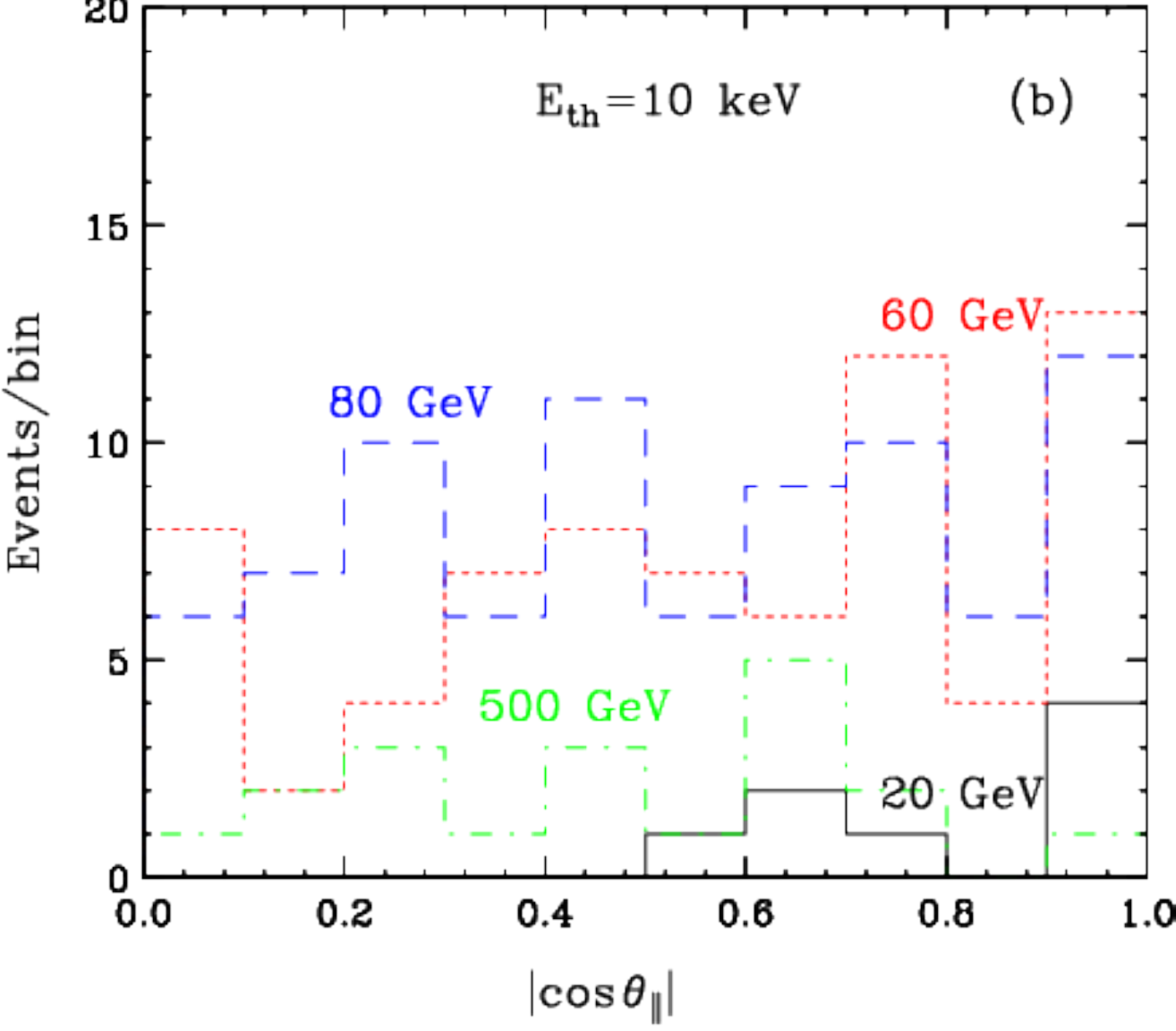}\hspace{0.07cm}
\includegraphics[scale=0.33]{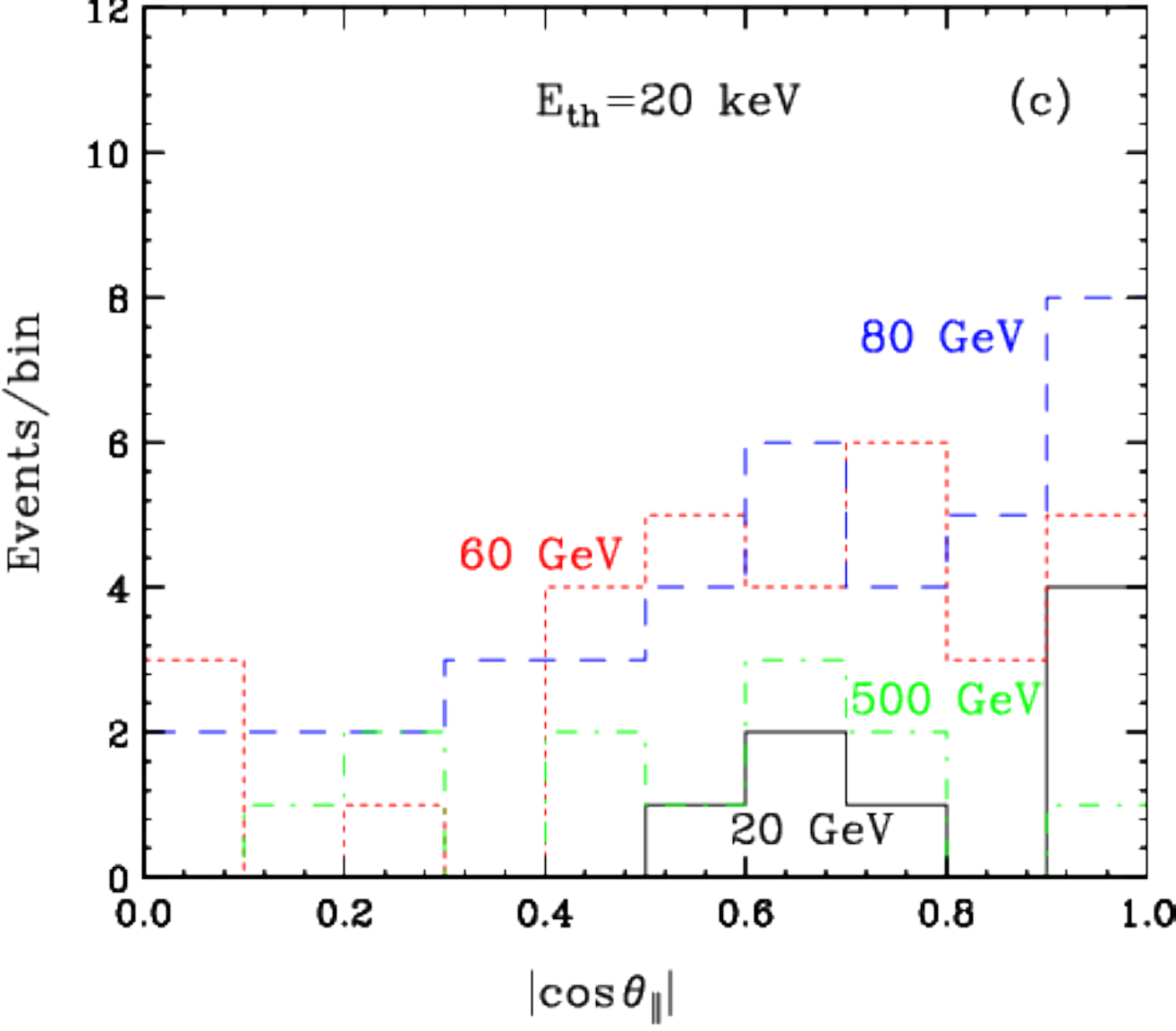}}
\caption{Effect of a threshold cut in angular distributions. 
4 keV (a), 10 keV (b) and 20 keV (c) energy threshold cuts are applied in each panel with a 20\% detection efficiency 
for 20 GeV (in black), 60 GeV (in red), 80 GeV (in blue) and 500 GeV (in green).}
\label{cosSIEth}
\end{figure}

We simulate the amount of scintillation light and ionization yield that would be obtained for several WIMP masses with a cross-section of $5 \times 10^{-11}$ pb in 10 ton-year exposure of a Xenon detector. Converting ($S$, $I$) to ($E_R$, $| \cos\theta_L |$), 
we obtain the results shown as the solid histogram in Figure \ref{cosSI} for two different $\epsilon$'s. The dotted histograms are illustrated in Figure \ref{para41020}, generated from an MC based on the theoretical expectations, 
assuming Gaussian smearing with 4 keV energy threshold. It is crucial to achieve a high photon detection efficiency as shown in Figure \ref{cosSI}(b), which shows a close match with results (dotted histogram in Figure \ref{cosSI}) in Figure \ref{para41020}.
However at a dark matter mass of 80 GeV, the $| \cos \theta_\parallel|$ distribution becomes flat even with a 10 ton-year exposure. The peak at $\cos\theta_\parallel = 0$ is a result of the very low light signal expected in these conditions and Poisson fluctuations shifting events towards $S=0$, hence $\cos\theta_\parallel = 0$.
It gets reduced for a better efficiency and/or a higher threshold cut as shown in Figures \ref{cosSI} and \ref{cosSIEth}. With a higher energy threshold cut, hence higher photo-statistics, the correct form of the angular distribution is recovered.

\begin{figure}[t]
\centerline{
\includegraphics[scale=0.33]{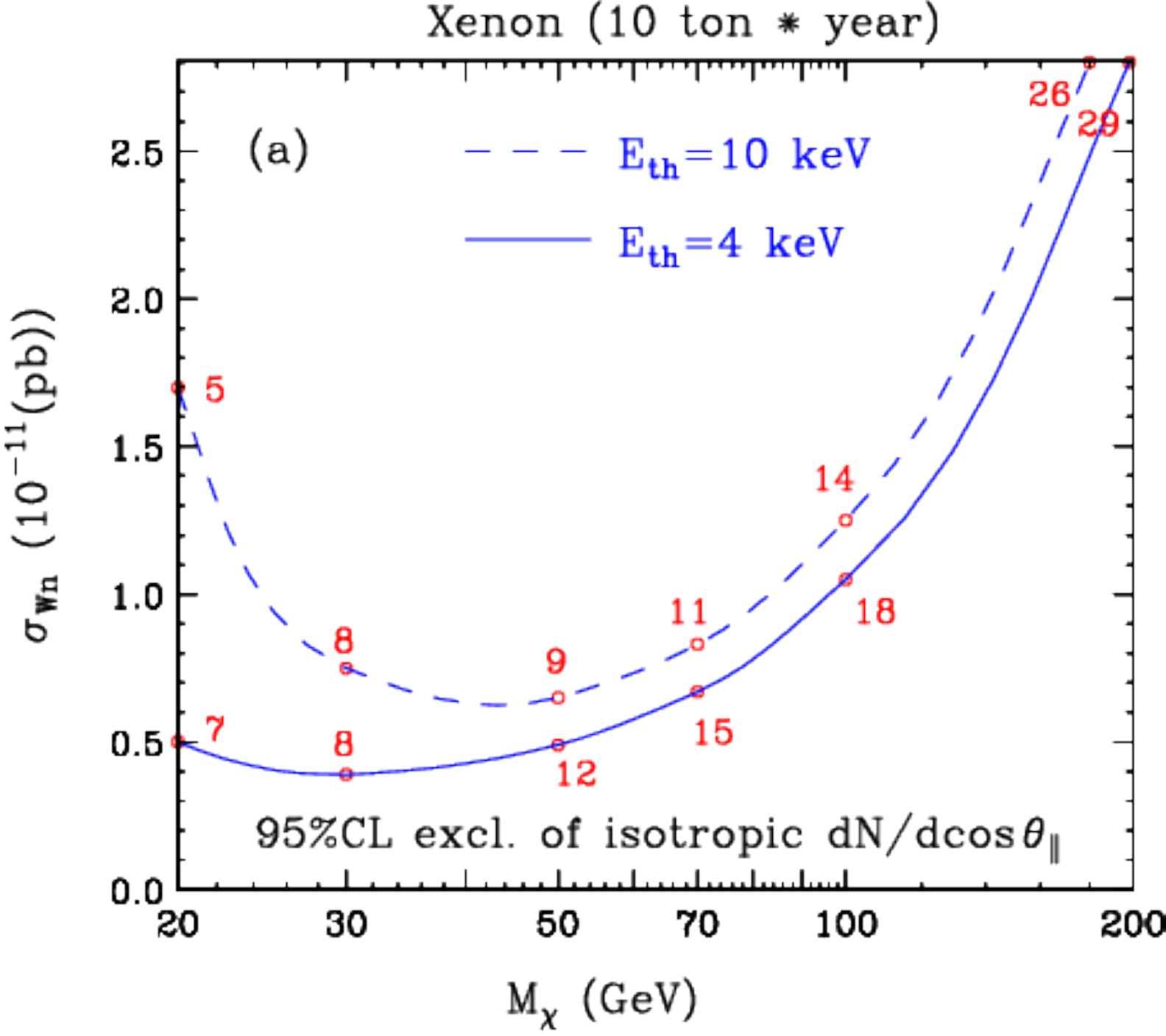} \hspace{-0.04cm}
\includegraphics[scale=0.33]{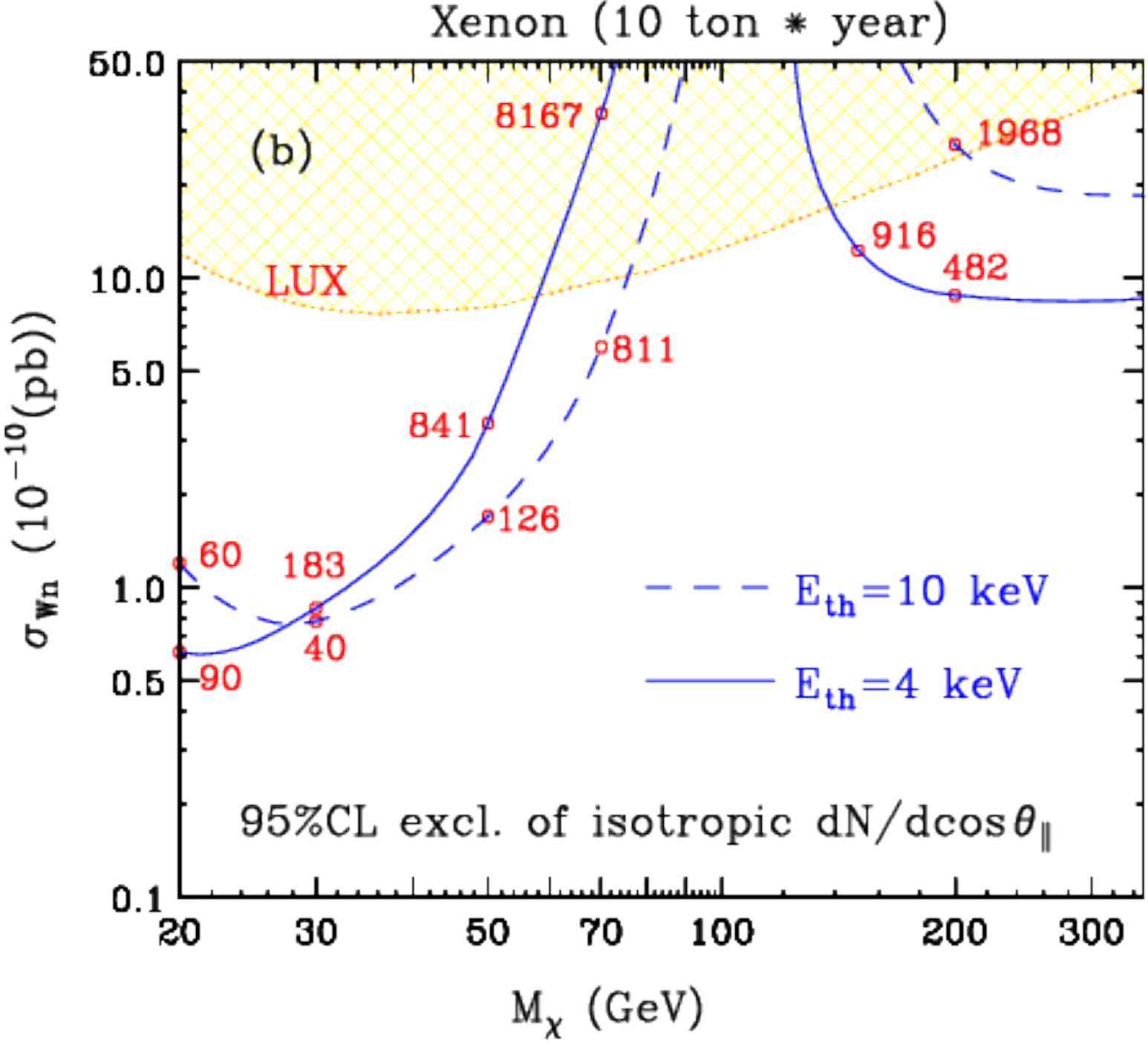} \hspace{-0.04cm}
\includegraphics[scale=0.33]{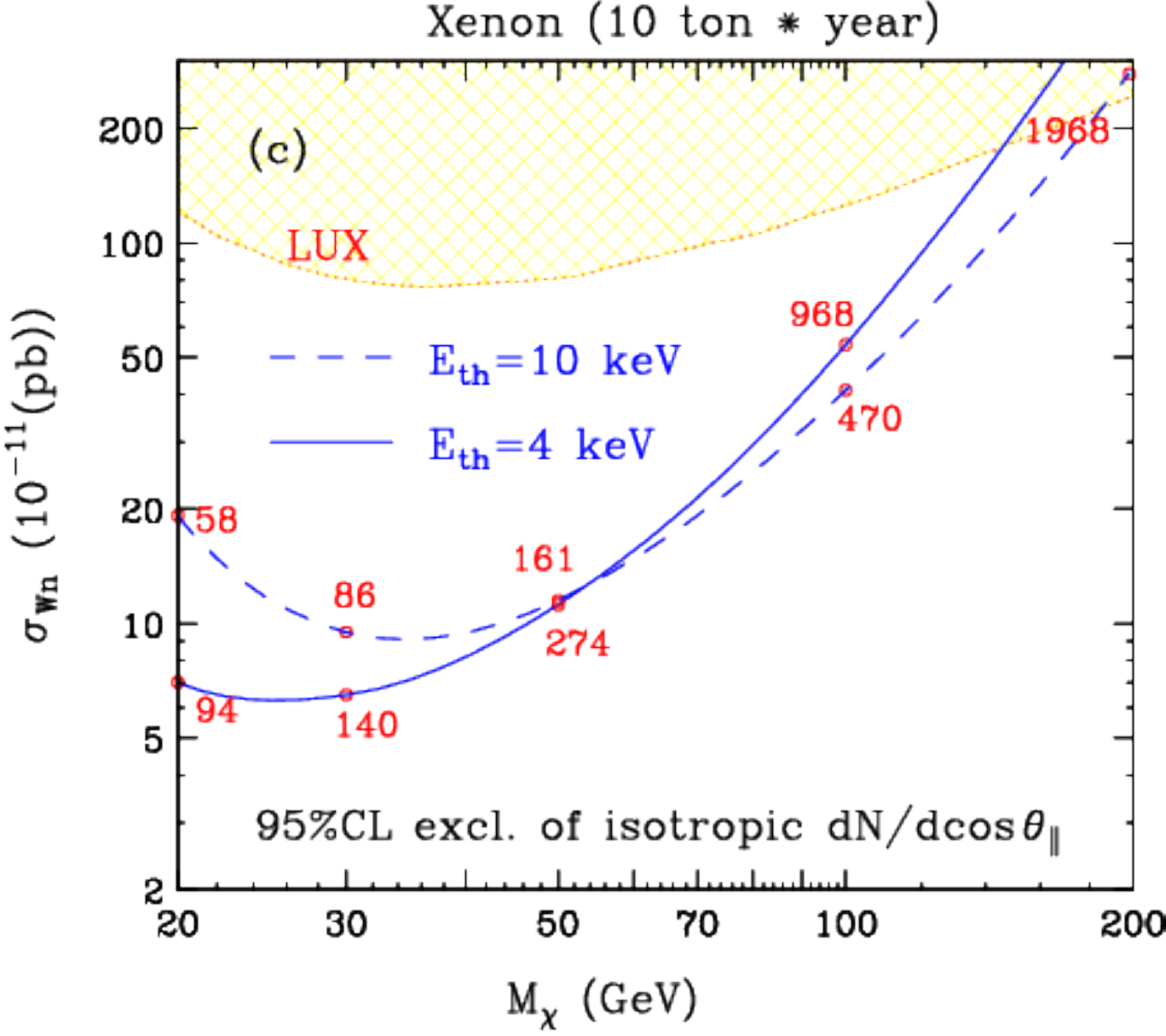} }
\caption{Test of the anisotropy for 
(a) head-to-tail capability, 
(b) no head-to-tail case, and
(c) the case with two angular bins. 
The solid (dashed) curves represent the boundary of exclusion of a flat angular distribution at 95\% CL with 4 keV (10 keV) energy threshold. 
The numbers (in red) provide the estimated number of events required to exclude a flat angular distribution, i.e., for directionality at a certain WIMP mass. An electric field parallel to the recoil trajectory is considered for 10 ton-year high-pressure Xenon detector exposure. The shaded area represents current exclusion limit by LUX \cite{Akerib:2013tjd}.
}
\label{anisotropy}
\end{figure}

We perform a likelihood ratio test to compute the required number of events to rule out the hypothesis of an isotropic velocity distribution. 
The results are shown at 95\% CL in Figure \ref{anisotropy}. 
The test has been done for three cases: 
(a) head-to-tail case,
(b) no head-to-tail capability, 
(c) the case with two angular bins.
Solid (dashed) curve represents the exclusion limit (at 95\% CL) that rules out hypothesis of a flat angular distribution 
(isotropic dark matter distribution) with 4 keV (10 keV) energy threshold. 
In other words, one can expect to observe anisotropy of dark matter distribution in the parameter space above the exclusion curve. 
The numbers in red give the required number of events needed at a certain WIMP mass to rule out an isotropic dark matter flow for 10 ton-year high-pressure Xenon detector exposure. Results are presented for an electric field parallel to the direction of the WIMP particles.
We have imposed $30^\circ$ angular resolution. The results are obtained with a theoretical double differential distribution, consistent with the dotted histogram in Figure \ref{cosSI}.
The shaded area represents the current exclusion limit by LUX \cite{Akerib:2013tjd} after an appropriate rescaling of 0.0275 ton-year exposure to our case
(0.0275 ton-year = 118 kg $\times$ 85 days).

\section{Discussion and Outlook\label{sec:conclusion}}
An observation of the anisotropy of dark matter interactions would provide decisive evidence for the discovery of dark matter \cite{Lee:2014cpa,Alves:2012ay,Kavanagh:2015aqa,Green:2010gw}. We investigated the feasibility of a high-pressure Xenon TPC dark matter detector which is sensitive to the angles of recoil produced in the interaction of dark matter particles with nuclei in the detector. The angular information helps precision measurements in the parameter space of the cross-section vs WIMP mass. Our study shows that full angular coverage and directionality could significantly improve the precision of the determination of the dark matter mass and/or the interaction cross section, especially for a heavy dark matter. The improvement is marginal for a detector without head-to-tail information or for a detector with the 2 angle bins. We find also that angular resolution does not make much difference in the improvement of the DM signal. 

The angular information of the recoil trajectory can be used to establish the anisotropy of the observed signal. A Xenon detector with a 10 ton-year exposure with head-tail capabilities could demonstrate the anisotropy of an observed signal if the corresponding interaction cross section is of the order of tens of $10^{-11}$ pb. The sensitivity of a detector without head-tail capabilities would be reduced by at least one order of magnitude.
It is interesting to notice that precision measurement is sensitive to a heavy dark matter while the
anisotropy probe is more sensitive to a light dark matter for this type of a high pressure Xenon detector.

\acknowledgments
We thank D. Nygren for useful the discussion and comments, and 
Azriel Goldschmidt for pointing out an issue in our normalization.
GM is partially supported by the National Research Foundation of South Africa under Grant No. 88614.
GM and KK are supported partially by the U.S. DOE under Grant No. DE-FG02-12ER41809 and by the University of Kansas General Research Fund allocation 2301566. Fermilab is operated by Fermi Research Alliance, LLC, under Contract DE-AC02-07CH11359 with the United States Department of Energy.


\bibliographystyle{JHEP}
\bibliography{refs}

\end{document}